\documentclass[sigconf,nonacm]{acmart}

\usepackage{textcomp}
\usepackage{etoolbox}
\robustify\textsc
\usepackage{listings}
\usepackage{xcolor}
\lstnewenvironment{PromptBox}{
  \lstset{
    basicstyle=\ttfamily\small, 
    frame=single,              
    framerule=0.5pt,           
    breaklines=true,           
    columns=fullflexible,      
    keepspaces=true,           
    showstringspaces=false     
  }
}{}
\usepackage{microtype}
\AtBeginDocument{%
  }

\usepackage{enumitem}
\usepackage{hyperref}

\begin{document}

\title{LumiNote: LLM-Assisted Multimodal Instruction for VR Stage Lighting Education}


\author{Danxuan Liang}
\orcid{0009-0008-4449-7764}
\email{dliangac@connect.ust.hk}
\affiliation{%
 \institution{The Hong Kong University of Science and Technology}
 \city{Hong Kong SAR}
 \country{China}
}

\author{Chun Yin Li}
\orcid{0009-0002-1408-9993}
\email{cylibl@connect.ust.hk}
\affiliation{%
 \institution{The Hong Kong University of Science and Technology}
 \city{Hong Kong SAR}
 \country{China}
}

\author{Zheng Wei}
\orcid{0000-0001-7444-2547}
\email{zhengwei@kaist.ac.kr}
\affiliation{%
 \institution{Korea Advanced Institute of Science and Technology}
 \city{Daejeon}
 \country{Republic of Korea}
}

\author{Xian Xu}
\orcid{0000-0002-2636-7498}
\email{xianxu@ln.edu.hk}
\affiliation{%
 \institution{Lingnan University}
 \city{Hong Kong SAR}
 \country{China}
}

\author{Meng Xia}
\orcid{0000-0002-2676-9032}
\email{mengxia@tamu.edu}
\affiliation{%
 \institution{Texas A\&M University}
 \city{College Station}
 \country{USA}
}

\author{Huamin Qu}
\orcid{0000-0002-3344-9694}
\email{huamin@cse.ust.hk}
\affiliation{%
 \institution{The Hong Kong University of Science and Technology}
 \city{Hong Kong SAR}
 \country{China}
}

\author{Wai Tong}
\orcid{0000-0001-9235-6095}
\email{wtong@tamu.edu}
\affiliation{%
 \institution{Texas A\&M University}
 \city{College Station}
 \country{USA}
}
\authornote{Corresponding author}

\renewcommand{\shortauthors}{Liang, D. et al.}


\begin{abstract}
Stage lighting education requires instructors to bridge abstract concepts, technical operations, and learner-understandable representations. While Virtual Reality (VR) removes physical constraints, existing systems provide limited support for live instruction. We present LumiNote, an LLM-assisted VR system that transforms spoken pedagogical intent into instructor-reviewable spatial annotations, executable demonstrations, and linguistic support. In an exploratory study with 3 instructors and 24 students, we examined how instructors incorporated LumiNote into familiar lighting topics and how students received the resulting representations. We found LLM assistance most valuable for expressive, under-specified goals, but requiring greater expert intervention for fixture-specific or spatial configuration requests. Instructors engaged with generated suggestions as a controllable refinement process, shifting effort from manual setup toward pedagogical expression. However, representations that externalized expert reasoning did not always align with novice comprehension. These findings characterize LLM-assisted VR instruction as a domain-grounded mediation process among expert expression, executable operations, and learner-facing representations.

\end{abstract}





\begin{teaserfigure}
    \centering
    \includegraphics[width=1\textwidth]{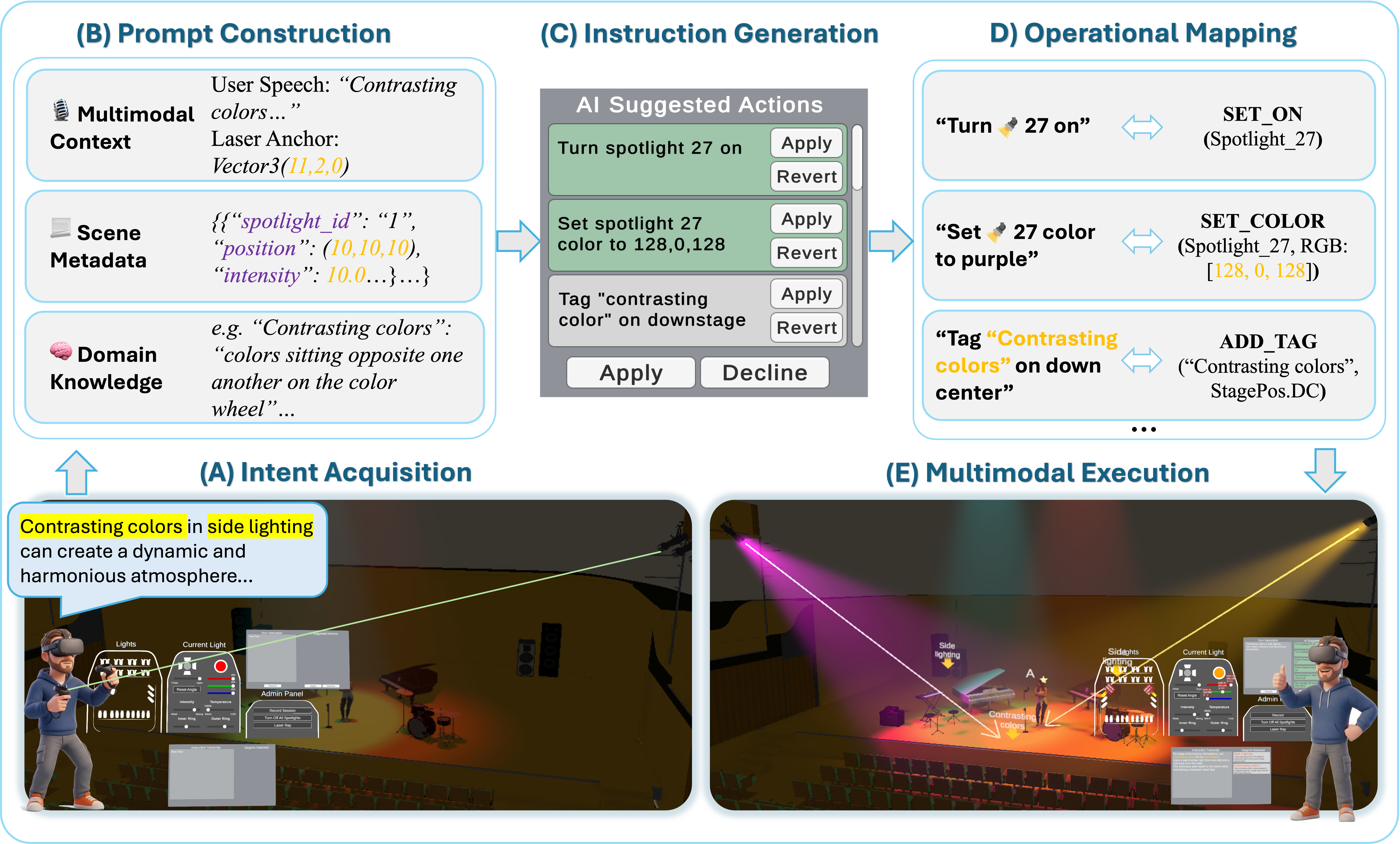}
    \caption{We propose LumiNote, an LLM-assisted system that transforms an instructor's high-level instructional intent during stage lighting education into multimodal instruction in virtual reality. (A) Intent Acquisition: An instructor provides voice commands (e.g., ``contrasting colors...'') while using a laser pointer to anchor the context to specific fixtures. (B) Prompt Construction: The system fuses multimodal user input, real-time scene metadata, and domain knowledge of stage lighting into a structured prompt. (C) Instruction Generation: The LLM interprets the intent to generate a set of suggested lighting demos and spatial annotations actions. (D) Operational Mapping: Suggested actions are parsed into executable API calls. (E) Multimodal Execution: The virtual theater updates instantly, synchronizing dynamic lighting effects with spatial annotations to reinforce instruction. All generated actions remain reviewable and are not executed without instructor approval.}
    \label{fig:overview}
\end{teaserfigure}

\maketitle
\section{Introduction}
Technical theater arts combine artistic expression with complex technical operations, requiring instruction that tightly integrates conceptual explanation with hands-on demonstration~\cite{essig_stanley_2007}. In stage lighting education, instructors routinely move among aesthetic intent, fixture-level decisions, visible stage effects, and learner-facing explanations. While Virtual Reality (VR) offers transformative potential by providing risk-free, accessible training environments, existing research has primarily focused on simulating lighting systems rather than supporting instructors' live teaching work~\cite{wei_feeling_2023,wei_illuminating_2025}. Consequently, within VR teaching environments, instructors face significant ``instructional friction'' as they coordinate spatial manipulations, technical configurations, and in-situ explanations. This creates a recurring translation challenge: instructors must translate pedagogical intent into concrete scene manipulations while simultaneously making those manipulations understandable to learners. Existing immersive systems provide limited support for either translation, leaving experts to bridge the gap between instructional intent and executable actions, and between technical demonstrations and learner-understandable representations.

To move beyond simulation toward instruction-centered VR, we conducted a formative study with four experienced stage lighting instructors to identify pedagogical needs in VR environments. We derived three design requirements: (1) situated spatial annotation that make abstract spatial relationships visible and intuitive, (2) on-demand real-time lighting demonstrations to maintain instructional flow without manual setup delays, and (3) linguistic support  to help instructors explain domain-specific terminology. These requirements highlight three instructor-facing mediation needs: spatializing explanation, accelerating demonstration, and translating domain language into learner-accessible representations.

Guided by these requirements, we developed \textit{LumiNote}, an instructor-facing LLM-assisted VR system that mediates between instructor intent, executable demonstrations, and learner-facing representations (Figure 1). During instruction, an instructor can express pedagogical intent through natural speech while grounding it in the 3D environment (e.g., ``use contrasting colors here''). LumiNote combines this multimodal input with scene metadata and domain knowledge to generate executable lighting actions and corresponding instructional representations. For instance, LumiNote can suggest spotlight-parameter adjustments to create contrast, overlay directional cues to visualize beam behavior, and tag relevant fixtures to connect terminology with physical elements. These outputs are mapped to executable operations and presented as reviewable suggestions before being applied in the virtual theater, allowing instructors to demonstrate abstract concepts while maintaining instructional flow and pedagogical control.

We conducted a two-phase exploratory evaluation: instructor-side use and learner-side reception. Three instructors who had not participated in the formative study taught the same predefined lighting topics first in a \textit{No LLM} VR session and then with LumiNote, allowing us to examine how LLM assistance was incorporated into familiar teaching workflows. Twenty-four students were randomly assigned to experience either a \textit{No LLM} or a \textit{With LLM} VR teaching session created by a single instructor, allowing us to compare learner reception while controlling for instructor style and pacing. We drew primarily on instructors' interviews and observed teaching behaviors, complemented by interaction logs and questionnaires to triangulate emerging patterns, while student responses and task outcomes provided a learner-side perspective on the resulting instructional content.


The study surfaced several broader patterns in LLM-assisted VR instruction. First, the value of LLM assistance varied with instructional intent: it was particularly useful for under-specified, expressive lighting goals, whereas precise fixture-level requests required greater expert intervention. Second, LLM assistance functioned as a controllable refinement process, with instructors retaining authority over how generated suggestions were accepted, rejected, or iteratively refined within their teaching flow. Third, LLM assistance shifted effort from manual demonstration setup and parameter adjustment toward pedagogical expression. Finally, instructor and learner evidence revealed a representation-alignment tension: representations that helped instructors externalize expert reasoning were not always those that novices found easiest to follow. Supporting measures contextualized these patterns: instructor workload and session duration were descriptively lower with LLM assistance, student presence was higher, and immediate task performance was comparable across conditions.





The primary contributions of this work are:

\begin{itemize}
\item An empirical characterization of instructor-facing mediation needs in VR stage-lighting instruction, identifying recurring needs for spatial explanation, rapid demonstration, and terminology mediation.

\item \textit{LumiNote}, an instructor-facing LLM-assisted VR system that translates under-specified pedagogical intent into constrained, reviewable scene actions and learner-facing representations.

\item An exploratory account of LLM-assisted VR instruction showing that the value of generative assistance varies with instructional intent, that instructors use generated suggestions as a controllable refinement process, and that expert-facing representations do not always align with learner-facing needs.

\end{itemize}

\section{Related Work}

\subsection{Immersive Systems for Learning and Training}
VR technology has gained significant popularity in professional skills training today, as it alleviates traditional resource constraints and offers a higher sense of presence for practical skill training than 3D desktop applications \cite{shu_virtual_2019}. Examples include viticulture skills training \cite{bondesan_implicit_2025}, geological surveying skills \cite{young_picks_2025}, and even fine skills such as medical device operation \cite{enderling_enabling_2025, wang_explainmr_2025}, or standardized laboratory procedures \cite{pieschacon_smart_2025, xue_artifacts_2023}. Recent VR instructional systems have also explored guided practice and procedural learning. For example, LearnIoTVR supports students in acquiring IoT concepts through interactive VR practice \cite{zhu2023learniotvr}. Beyond VR, immersive instructional systems have also explored tutorial authoring; for example, InstruMentAR automatically generates step-by-step AR tutorials for operating digital instruments from recorded embodied demonstrations \cite{liu2023instrumentar}. 
Together, these systems demonstrate how immersive environments can support structured practice, procedural learning, and tutorial-based instruction.

Similar attempts have also been made in technical arts, such as VR systems for film shooting training \cite{wang_vaction_2025}. In addition to pre-designed in-VR tutorials, research has also explored how collaborative metaverse environments can support professional skill development. For example, interactive training systems for film and stage lighting design \cite{wei_feeling_2023, tong_exploring_2024,wei2025exploring} allow instructors and students to interact, learn, and practice together in virtual spaces. Other work has investigated how single‑user versus multi‑user collaboration in different virtual environments influences learning outcomes in cinematography lighting \cite{wei_illuminating_2025}.

However, existing immersive learning systems primarily support structured practice, simulation, or authored tutorials, with comparatively limited attention to instructors’ spontaneous work during live teaching. Although immersive environments can externalize difficult-to-observe information such as light paths and spatial relationships and couple these representations with ongoing demonstrations, few systems integrate such capabilities into instructor-led pedagogical workflows.

\subsection{
Stage Lighting Pedagogy and Digital Tools
}
Stage lighting education relies heavily on instructors' real-time guidance and demonstration \cite{essig_stanley_2007}. During instruction, instructors must move among expressive concepts, spatial relationships, fixture-level operations, and visible lighting effects while making these connections understandable to learners. Beyond immersive training systems, stage lighting education also draws on professional lighting previsualization and design tools, such as \textit{Vectorworks Spotlight} \cite{noauthor_stage_nodate}, \textit{Capture} \cite{noauthor_capture_nodate}, and \textit{WYSIWYG} \cite{noauthor_wysiwyg_nodate}. These tools reproduce lighting fixtures, visual effects, and operational workflows and can reduce dependence on access to physical venues and equipment. However, they primarily support lighting design and simulation rather than the instructional work of connecting technical operations with spatial behavior, perceptual effects, and design intent.

In practice, instructors use several complementary approaches to support concept explanation and demonstration. Slides and textbook examples allow instructors to use lines \cite{matt_kizer_scenic_2020, aputure_lighting_2020}, arrows \cite{matt_kizer_scenic_2020, alan_hamilton_audio_mastering_2024}, and other annotations to explain lighting concepts and operations \cite{thanasoula_learning_2024}. However, their 2D and pre-authored nature limits embodied spatial demonstration and real-time manipulation. Recorded classroom demonstrations can capture the use of physical equipment and support later explanation \cite{baecher_supervisor_nodate}, but they separate instructional feedback from the moment in which lighting decisions and adjustments are made \cite{hattie_power_2007}. Interactive simulation tools support real-time manipulation of lighting effects, yet instructors often need additional interfaces or annotation tools, such as screen projection or third-party software, to connect visible changes with conceptual explanations and operational details. Prior work on one-to-many AR instruction further suggests that coordinating guidance across multiple learners places additional demands on instructor capacity \cite{otsuki_assessment_2022}.

Taken together, existing tools support individual parts of stage lighting instruction, including conceptual explanation, recorded demonstration, and interactive operation, but leave instructors to manually connect aesthetic intent, spatial relationships, fixture-level actions, and learner-facing representations. This motivates instructional support that can connect these layers within a shared spatial context while preserving the flow of ongoing teaching.

\subsection{LLM-Assisted Teaching and Educational Workflows}
Recent LLM-assisted educational systems support a range of instructional activities, including lesson preparation, assessment, knowledge support, and personalized coaching. Instructor-facing systems have assisted with tasks such as lesson preparation \cite{kang_tutorcraftease_2025} and collaborative report grading \cite{chen_cograder_2025}. Other systems provide learner-facing support during learning, including real-time jargon explanations \cite{liu_exploring_2025}, domain-specific practice guidance \cite{blanchet_integrating_2023}, and personalized coaching feedback, such as AgentCoach \cite{ma2026agentcoach}. Multimodal models have also been explored for automated evaluation and generative feedback across creative domains, including UI design \cite{duan_uicrit_2024}, painting creation \cite{zheng_artmentor_2025}, and filmmaking \cite{darejeh_filmmaking_2025}.

While these systems demonstrate the potential of LLMs across educational workflows, comparatively little work has examined how LLM assistance can support instructors during live pedagogical delivery. In such settings, instructors must simultaneously articulate pedagogical intent, coordinate demonstrations, and communicate concepts to learners while adapting to the ongoing teaching context. This form of support involves two linked translations: from an instructor's high-level pedagogical intent to a concrete instructional demonstration, and from that demonstration to a representation that learners can understand. For example, an instruction such as ``make the performer stand out with warmer light'' must first be realized through concrete lighting changes, while those changes may also need to be accompanied by spatial annotations or technical explanations to make their pedagogical meaning clear to learners. Our work investigates LLM assistance as an instructor-facing mediation mechanism that connects pedagogical intent, instructional demonstrations, and learner-facing representations during live VR teaching.
\subsection{Language-Based XR Grounding and Authoring}

Recent work has explored how generative AI can interpret natural language together with spatial input to support interaction in immersive environments. One line of work focuses on multimodal spatial intent grounding, where LLM-based systems combine language with embodied cues to resolve references to objects, locations, and spatial relationships in 3D environments. For example, GesPrompt parses co-speech gestures to extract spatial-temporal parameters such as position, direction, scale, and motion paths \cite{hu_gesprompt_2025}; VRMover combines speech and pointing for natural multi-object manipulation in VR \cite{wang_can_2025}; and GazePointAR uses gaze and pointing cues to resolve deictic references in spatial commands \cite{lee2024gazepointar}. Together, these systems demonstrate how multimodal interaction can help ground underspecified language in the objects, locations, and spatial relationships of an immersive environment.

A related line of work explores language-based XR authoring and content generation. LLMR \cite{de2024llmr} enables the real-time creation and modification of interactive 3D worlds through language-based interaction, while DreamCodeVR supports voice-driven programming of behaviors for 3D content \cite{giunchi2024dreamcodevr}. VRCopilot assists users in specifying immersive layouts through speech and pointing \cite{zhang2024vrcopilot}, and AgentAR \cite{zhu2025agentar} uses tool-augmented agents to support the authoring of AR applications. Beyond basic prompt execution, systems like SceneCraft \cite{hu2024scenecraft} and LLMER \cite{chen2025llmer} translate language into structured code and scene graphs. More recently, HOICraft \cite{lee2026hoicraft} and Roomify \cite{wang2026roomify} extend language-driven intent to part-level interaction design and spatially grounded scene transformation. Together, these systems show how LLMs can translate high-level user intent into structured operations for generating, arranging, and modifying XR content.

However, this body of work primarily focuses on general-purpose object manipulation, world building, layout authoring, and application creation. Comparatively little work has examined pedagogically situated grounding during live instruction, where an instructor's high-level intent must be translated into executable actions while preserving its instructional meaning and supporting learner understanding. In this setting, technical executability alone is insufficient because generated actions must remain inspectable and controllable by the instructor and may also need to be expressed through representations that learners can follow. LumiNote investigates this form of pedagogically situated grounding by connecting instructors' spoken intent with bounded, reviewable scene actions and corresponding learner-facing representations during live VR instruction.

\section{Formative Study}

To understand the pedagogical needs of stage lighting instructors, we conducted semi-structured interviews with four experts (P1–P4). The group included two stage lighting instructors (5 and 3.5 years of experience) and two theater arts professors (15 and 7 years of experience). Each interview lasted approximately 60 minutes and was conducted via video conferencing. The goal of the discussion was to determine how an immersive system could better support real-time instruction for stage lighting educators.

\subsection{Protocol}
The semi-structured interviews followed a qualitative inquiry framework designed to capture existing pedagogical workflows and identify the requirements for immersive instruction. Each session was organized into three areas: \textbf{current workflows and pedagogical challenges}, \textbf{factors influencing system adoption}, and \textbf{design reflection and pedagogical utility}.
First, we asked questions about current workflows and pedagogical challenges (e.g., could you outline the typical process for instructing students in a stage lighting course?). Participants described their instructional environments, including the physical consoles or digital platforms used to demonstrate lighting concepts. We focused on the translation gap between abstract artistic intent and technical execution while exploring barriers to adopting new technologies. Second, we are interested in questions regarding factors influencing system adoption (e.g., what is your preference regarding providing instruction through a VR interface versus your traditional tools?). We investigated expert preferences for providing instruction through a VR interface versus traditional tools and identified the primary motivations for adopting immersive tools.
Finally, during design reflection and pedagogical utility, we presented our initial VR system concept (simulation only)
and asked experts to evaluate how multimodal interactions could address their current instructional barriers (e.g., what features do you consider most valuable in an instructor interface specifically for the Stage Lighting domain within this VR system?). All interview questions can be found in the supplementary materials.

\subsection{Interview Results and Findings}
We transcribed the interviews and conducted a thematic analysis to identify recurring challenges and functional requirements in stage lighting pedagogy. A researcher first segmented the transcripts according to the interview structure, including current feedback practices, existing tools, teaching workflows, perceived VR affordances, desired instructor-interface features, and challenges in current instruction. Following this initial organization, two researchers independently coded the materials using an inductive approach. 

The final codebook captured both current instructional practices and future design opportunities. Current-practice codes described how instructors provide feedback through face-to-face discussion, visualization platforms, LMS-based comments, and real-time lab or artistry activities. Workflow-related codes further characterized stage lighting education as progressing from basic concepts and theory to practical operation, artistry, and production. Design-opportunity codes captured desired support for realistic 3D venue simulation, real-time spatial annotation, voice-based tagging and transcription, instant visual changes, before--after comparisons, session recording, communication helpers for professional terminology, and instructor control over student feedback. These codes show that the target teaching context was not a single interface feature, but a broader live-instruction scenario involving spatial explanation, operational demonstration, and terminology mediation.

Inter-rater reliability reached a Cohen's Kappa of 0.8927, indicating high agreement. Any discrepancies were resolved through iterative discussion to reach a final consensus. The final formative study codebook is provided in Appendix~\ref{app:formative_codebook}. We report these themes and the resulting design potentials as findings below.

\subsubsection{Current Pedagogical Workflow and Digital Gaps}
Instructors primarily use 2D pre-visualization tools, LMS-based feedback, and face-to-face discussion to support stage lighting instruction (P1, P2, P3). These tools are useful because physical lighting equipment and rehearsal venues are often scarce, but they also create a disconnect in real-time instruction. Current methods often rely on screen-sharing, post-session notes, or verbal explanation, which makes it difficult for students to immediately connect a demonstrated effect with their own practice (P2, P3). 

The interviews also showed that beginner-stage lighting education spans multiple learning goals, including fixture usage, focusing and plotting, spatial understanding of the venue, parameter adjustment, visual effects, and production-oriented composition. However, 2D tools cannot fully reproduce the spatial scale and embodied viewpoint needed to judge lighting in a real venue (P3, P4). This lack of spatial context, combined with students' diverse backgrounds, makes it difficult to explain abstract concepts, professional terminology, and aesthetic intentions through verbal description alone.
    
\subsubsection{Factors Influencing VR Adoption}
Participants saw VR as valuable for beginner instruction because it can show realistic equipment, provide a safe space for experimentation, simulate a theater venue, and reduce dependence on live performers or scarce physical resources (P1, P2, P3, P4). These affordances are especially relevant for early-stage learning, where students need repeated opportunities to try lighting setups without safety risks or venue constraints.

At the same time, instructors emphasized that realism alone is insufficient. If VR only replicates physical scenarios, the operational complexity and physical discomfort may outweigh the benefits of immersion. Several participants, therefore, preferred mobile or desktop simulations for routine feedback because they are easier to access and operate. This suggests that VR must provide functional advantages beyond simulation, such as making invisible lighting relationships visible, accelerating live demonstrations, and supporting feedback that is difficult to deliver in physical or 2D settings.
    
\subsubsection{Exploring ``Beyond-Reality'' Potentials}
To address these barriers, we guided participants to envision VR features that provide ``beyond-reality'' support. Experts identified three high-value areas: \textbf{multimodal feedback}, \textbf{automated demonstrations}, and \textbf{linguistic assistance}.

They suggested that real-time in-scene tagging and circling of relevant regions would assist spontaneous explanations (P1, P2, P3, P4). For example,
\textit{``Tagging can point out tiny mistakes directly.''} (P1), \textit{``If I can annotate within the 3D environment to identify an instrument like ``change this color'' or ``swap this lens,'' it really helps.''} (P3) and \textit{``It's better to combine visual annotations with verbal comments for real-time instruction.''} (P4)

To reduce the operational burden, they expressed a need for the system to generate lighting effects instantly based on intent (P1, P2, P3). 
For example, \textit{``If I can circle the specific area I am discussing and instantly replace that light to see the difference in real time, it would be extremely helpful.''} (P2) and \textit{``Changing a lens from $25^\circ$ to $50^\circ$ takes time... if the system can capture my intent when I am talking about shades or adjusting angles and let me see the impact within that timeframe, it's really helpful.''} (P3)

Finally, P3 and P4 highlighted the need for communication helpers to bridge the gap between technical jargon and visual understanding.
For instance, \textit{``It gets complicated when I am talking about shades—whether my student is understanding the same thing
I am imagining... terminology usage varies across regions.''} (P3) and
\textit{``Students can be at very different levels,
so trying to cater a single course to everyone in the room can be complicated.''} (P4)

\subsection{Design Requirements}

We translated the formative findings into design requirements by focusing on the live-instruction breakdowns that VR and LLMs could uniquely address. Although the interviews surfaced a broad range of needs, including rigging safety, cabling, venue simulation, collaboration, and session review, LumiNote focuses on the instructor-facing moment of live concept delivery. We therefore prioritized requirements that directly support spatial explanation, immediate visual demonstration, and terminology mediation during real-time teaching.

Based on this scope, we propose the following design requirements.

\textbf{\textit{DR1: Situated Spatial Annotation for Explanation.}}
The system should support instructors in creating in-scene annotations that make spatial relationships, relevant objects, and operational cues visible within the ongoing teaching context. Such annotations should help instructors externalize information that is difficult to convey through verbal explanation alone and connect abstract concepts to elements in the 3D environment.

\textbf{\textit{DR2: On-Demand Real-Time Lighting Demonstrations.}}
To reduce the operational effort required during live demonstration, the system should support the on-demand generation and execution of lighting demonstrations. This provides instructors with actionable references and case-based examples during live teaching without interrupting the instructional flow.

\textbf{\textit{DR3: Linguistic Support for Technical Communication.}}
To address the need for clarity across diverse backgrounds, the system should provide communication assistance that helps instructors convey domain-specific jargon and complex lighting concepts clearly and efficiently during instruction.



\subsection{Pedagogical Expression Strategies for Stage Lighting Instruction}
\begin{figure*}
    \centering
    \includegraphics[width=\linewidth]{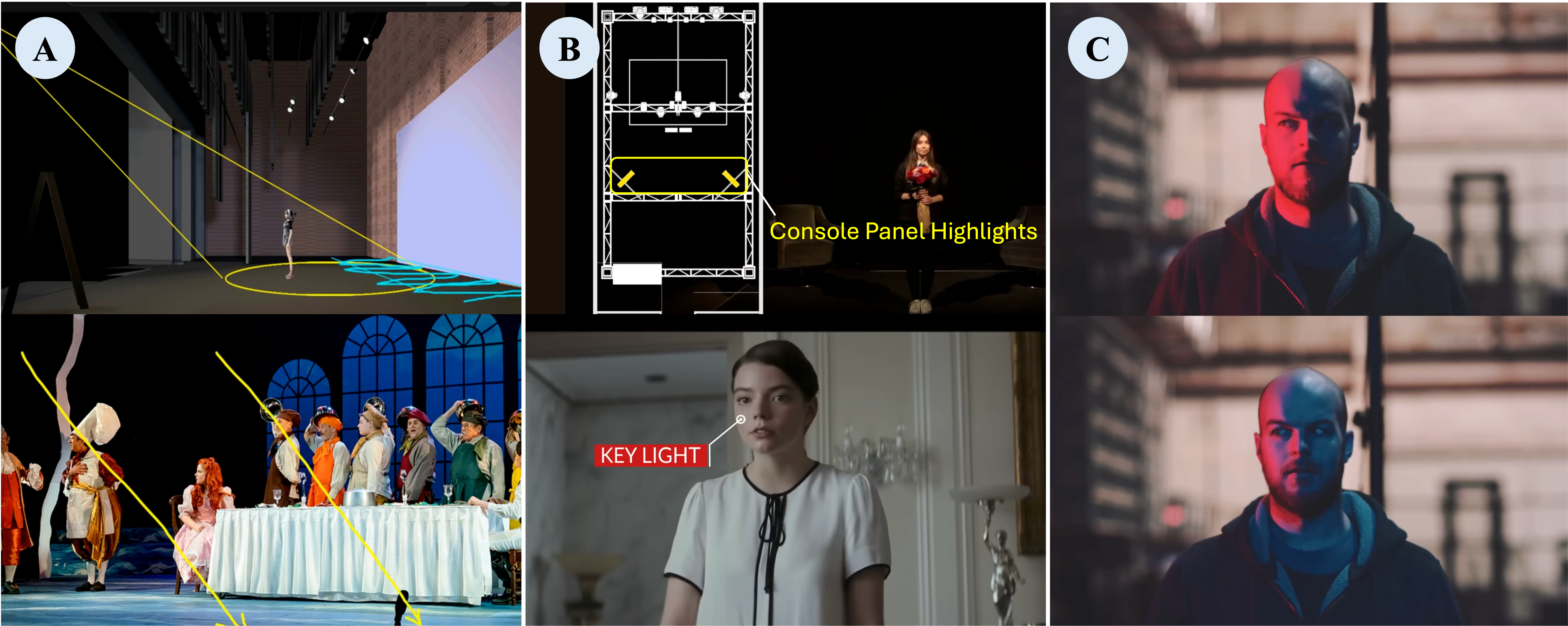}
    \caption{Representative pedagogical annotations. (A) Spatial mapping using 3D lines and arrows \cite{matt_kizer_scenic_2020}. (B) Contextual console panel highlights \cite{vari-lite_theatrical_2023} and tagging of scene objects \cite{aputure_lighting_2020}. (C) Comparison-based demonstration illustrating the visual shift between lighting concepts \cite{stage_right_basics_2017}.}
    \label{fig:annotation types}
\end{figure*}
To better support spatial annotations (\textbf{\textit{DR1}}) and real-time lighting demonstrations (\textbf{\textit{DR2}}), we analyzed five beginner-level stage lighting courses from five different lighting design content creators~\cite{matt_kizer_scenic_2020, aputure_lighting_2020, alan_hamilton_audio_mastering_2024, vari-lite_theatrical_2023, stage_right_basics_2017}. These videos are selected to cover foundational beginner topics in established stage‑lighting pedagogy~\cite{collins_lookingatlight}: \textbf{fixture and configuration}, \textbf{spatial behavior}, \textbf{visual effects}, and \textbf{composition strategies}.
From this corpus, we identified recurring pedagogical strategies for enhancing concept delivery and operational instruction. We selected strategies that addressed the formative design requirements, supported foundational lighting learning dimensions, and could be implemented in VR as instructor-triggered or LLM-generated representations. 

\begin{itemize}[leftmargin=*,topsep=0pt]
    \item \textbf{Spatial Geometry:} Spatially mapped arrows, text, and lines are used to emphasize directional vectors, light propagation, and area coverage (Figure \ref{fig:annotation types}A)~\cite{matt_kizer_scenic_2020, aputure_lighting_2020, alan_hamilton_audio_mastering_2024}. This strategy directly reinforces understanding of \textbf{spatial behavior} and \textbf{composition strategies} by helping learners visually connect abstract geometric concepts to physical stage lighting behavior.
    \item \textbf{Contextual Tagging:} Dynamic labels and visual highlights are overlaid on stage fixtures and console panels (Figure \ref{fig:annotation types}B) to link technical jargon with corresponding physical components and operations~\cite{matt_kizer_scenic_2020, aputure_lighting_2020, alan_hamilton_audio_mastering_2024, vari-lite_theatrical_2023}. This addresses \textbf{fixture and configuration} as well as \textbf{spatial behavior}, supporting situated, real-time operational understanding during instruction.
    \item \textbf{Actionable Demos:} Real-time demonstrations of lighting adjustments are employed (Figure \ref{fig:annotation types}C) to illustrate the effects of parameter changes~\cite{matt_kizer_scenic_2020, aputure_lighting_2020, stage_right_basics_2017}. This allows learners to observe how abstract controls translate into tangible \textbf{visual effects} and overall stage \textbf{composition strategies}.
\end{itemize}


\section{LumiNote}
\label{tab:luminote}
In this section, we present the design and architecture of LumiNote, an LLM-assisted VR system for stage lighting education, iteratively developed based on our distilled requirements. LumiNote interprets an instructor's intent from voice instruction and in-scene interactions to generate an instructor-reviewable list of actionable lighting adjustments and spatial annotations for concept-based demonstrations. It also transcribes the instructor's instruction, identifies and provides explanations of jargons in real-time. By turning spoken pedagogical intent into reviewable scene actions and representations, LumiNote helps instructors coordinate abstract explanation, technical operation, and learner-facing feedback within a shared immersive environment.

\subsection{System Design}
LumiNote is built within a foundational digital twin of a physical theater hall, featuring a 1:1 scale replication and a functional lighting grid that allows for manual operation via a controller-interactable console panel to support realistic pedagogical settings. LumiNote supports two distinct operational modes specifically tailored to the pedagogical roles of the instructor and the student, as shown in Figure \ref{fig:instructor&student}. While both users share a synchronized spatial context, their interactive capabilities are differentiated by role:

\begin{figure*}
    \centering
    \includegraphics[width=1\textwidth]{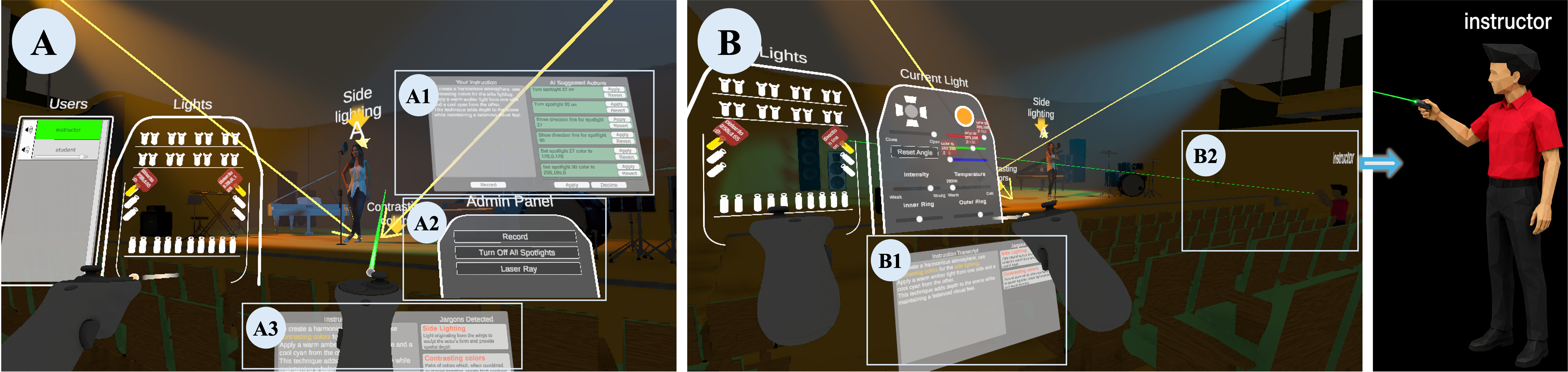}
    \caption{LumiNote User Interfaces. Instructor View (A): LLM interaction panel (A1), administrative toolkit (A2), and intelligent jargon panel (A3). Student View (B): Synchronized 3D stage and jargon panel (B1). Social Presence (B2): Instructor's avatar and broadcastable laser pointer from the student's perspective.}
    \label{fig:instructor&student}
\end{figure*}

\textbf{Instructor Mode (Figure \ref{fig:instructor&student}A):} The instructor is granted an authoritative toolkit for real-time content generation and session management. Beyond manual lighting operations, this includes exclusive access to the LLM Interaction Panel (Figure \ref{fig:instructor&student}A1) for generating actionable demos and annotations, an Intelligent Jargon Panel (Figure \ref{fig:instructor&student}A3) that transcribes instructions while highlighting and explaining domain jargon in real time, as well as an Administrative Toolkit (Figure \ref{fig:instructor&student}A2) for session recording, global resets, and broadcasting the 3D laser pointer. Further details on these elements are provided in Section \ref{tab: interaction design}.

\textbf{Student Mode (Figure \ref{fig:instructor&student}B):}
Designed for guided learning, students have access to manual lighting operations and a jargon panel for real-time review during instruction (Figure \ref{fig:instructor&student}B1). 
To ensure instructional consistency, students cannot trigger LLM actions; they observe expert-validated demonstrations and annotations only after the instructor formally applies them.
Real-time avatar synchronization (Figure \ref{fig:instructor&student}B2) maintains social presence and spatial awareness during the session.

\subsection{Instructor Mode Walkthrough}
During a live lighting lesson, the instructor can invoke LumiNote whenever they want to turn a spoken explanation into an immediate visual demonstration. As shown in Figure~\ref{fig:instructor interface}, the instructor first taps "Record'' in the LLM Interaction Panel (Figure~\ref{fig:instructor interface}A), speaks a teaching instruction, and stops the recording. For example, when explaining that ``contrasting colors can create a harmonic effect,'' the instructor simply delivers the spoken instruction naturally. LumiNote transcribes the speech, interprets the instructional intent, and presents a list of AI-suggested actions, such as changing spotlight colors, adjusting intensity, rotating a light toward a performer, showing a beam-direction arrow, or adding a concept tag in the scene.

Before any change is applied, the instructor can review the suggested actions and choose to apply, reject, or manually refine them through the lighting console (Figure~\ref{fig:instructor interface}B). When a lighting action is applied, LumiNote also adds console tags to the corresponding student-visible synchronized console panel, linking actions to light buttons (Figure~\ref{fig:instructor interface}B1) or parameter sliders (Figure~\ref{fig:instructor interface}B2), helping students connect the visual change on stage with the underlying lighting operation. At the same time, the transcribed instruction is displayed in the Instruction Transcript (Figure~\ref{fig:instructor interface}D1), where detected lighting jargon is highlighted and explained in the Jargon Panel (Figure~\ref{fig:instructor interface}D2). The instructor can further use the Admin Toolkit (Figure~\ref{fig:instructor interface}C) to record the session, reset all lights, or activate a laser ray for spatial pointing. Through this workflow, LumiNote supports instructors in coordinating speech, lighting actions, annotations, console feedback, and jargon explanations during live VR-based instruction.

\subsection{Interaction Design}
\label{tab: interaction design}
\begin{figure*}
    \centering
    \includegraphics[width=1\textwidth]{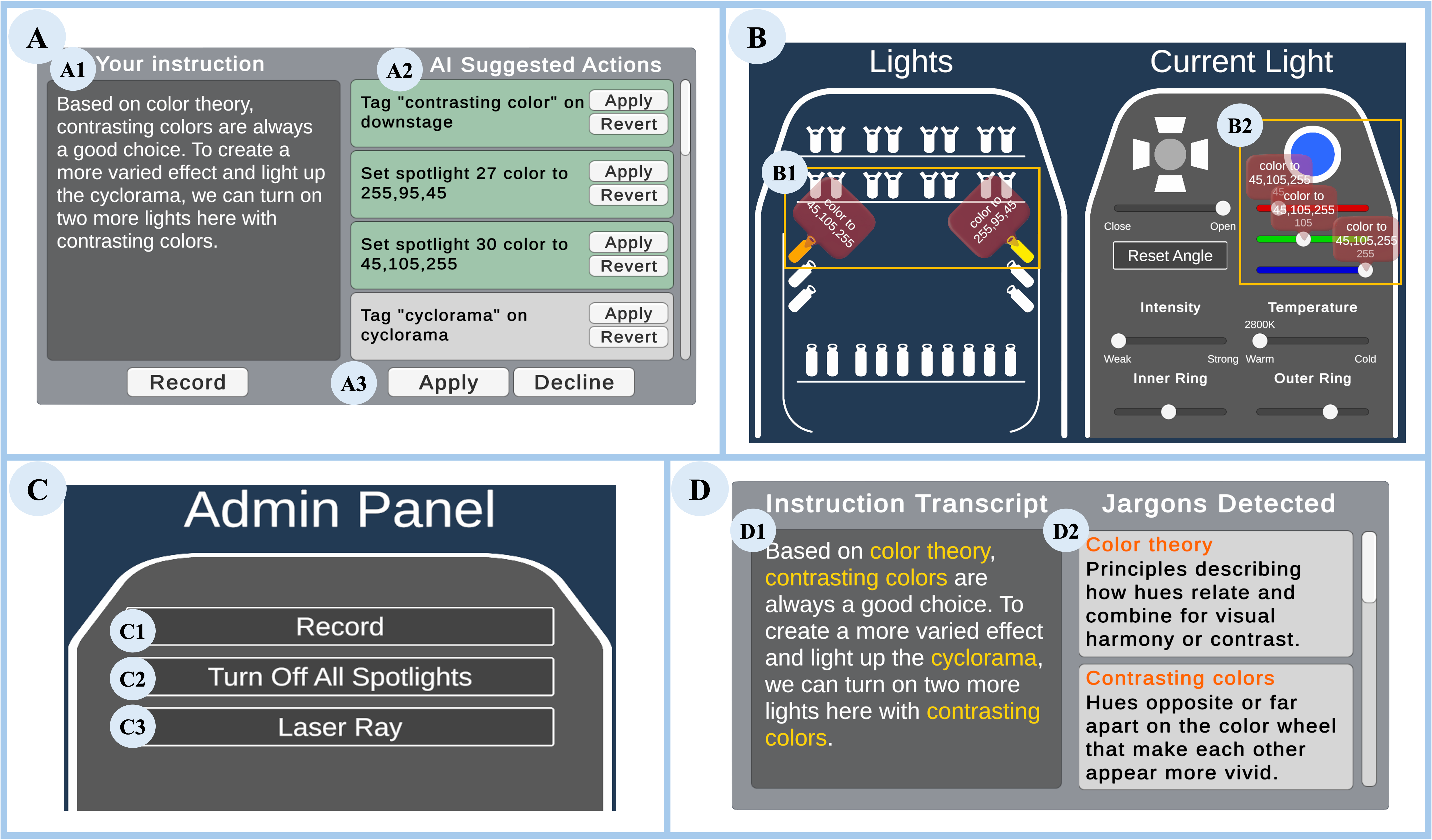}
    \caption{Instructor Interface Overview.(A) LLM Interaction Panel: (A1) instruction input, (A2) suggested actions with apply and revert options, and (A3) bulk controls to apply or decline all.(b) Lighting Console Panel: (B1, B2) control tags mapped to LLM actions.(C) Admin Toolkit: (C1) session recording, (C2) master blackout, and (C3) laser pointer.(D) Jargon Panel: (D1) transcript with highlighted technical terms and (D2) definitions for selected jargon.}
    \label{fig:instructor interface}
\end{figure*}

LumiNote features an integrated suite of interactive tools as shown in Figure \ref{fig:instructor interface} designed to facilitate real-time stage lighting instruction through intent-driven input, LLM-generated actionable suggestions, a synchronized jargon support system, and a manual administrative toolkit for global scene management.

\subsubsection{Intent-Driven Instruction Input}
To minimize cognitive load during live demonstrations, we employed a hands-free, voice-driven modality (Figure \ref{fig:instructor interface}A1). Instructors trigger audio recording via a single-tap controller interaction; the captured speech is then processed through the Whisper model and transcribed into real-time text for immediate interface verification.
\\
\subsubsection{Actionable Suggestion Types}
Once the instructional intent is parsed, the system generates three distinct types of visual and functional aids:

\textbf{Spatial Geometry:} The system renders 3D geometric primitives in virtual space to visualize invisible lighting properties, using spatial annotations to emphasize critical directional metrics (\textbf{\textit{DR1}}). This includes direction lines and beam vectors that trace the path from a spotlight to its target, helping to explain spatial coverage and beam angles.

\textbf{Contextual Tagging:} Dynamic annotations are applied directly to the 3D environment to link and explain jargon (\textbf{\textit{DR1, DR3}}). This includes placing semantic tags on specific fixtures (e.g., "Upstage", "Cyclorama") at their precise world positions and highlighting stage objects by a jumping star to clarify their relationship with the current lighting setup.

\textbf{Actionable Demos on Spotlights:} The system translates high-level aesthetic intent into executable DMX-based actions that directly manipulate the virtual fixtures in real-time (\textbf{\textit{DR2}}), while providing a synchronized visual guide on the control interface. These actions include:
\begin{itemize}[leftmargin=*,topsep=0pt]
    \item \textbf{Power Control:} Toggling spotlight power on/off and adjusting granular dimmer percentages.
    \item \textbf{Color Tuning:} Executing precise color shifts via specific RGB values and saturation levels.
    \item \textbf{Beam Control:} Modifying beam properties, including angles and the softness of the beam/field (inner/outer ring).
    \item \textbf{Spatial Targeting:} Rotating spotlights to align with a designated stage coordinate or target object.
\end{itemize}

To reinforce the learning outcome, the system generates Real-time \textbf{Console Tagging} for every adjustment (\textbf{\textit{DR1, DR3}}). These tags are pinned directly to the corresponding buttons (Figure \ref{fig:instructor interface}B1) and sliders  (Figure \ref{fig:instructor interface}B2) on the virtual console, explicitly labeling the operation (e.g., "RGB Color") and the exact target value (e.g., <255, 192, 203>). This creates a direct pedagogical link between the visual effect on stage and the technical input required on a professional lighting desk.

\subsubsection{Interaction with LLM-Generated Content}
To maintain pedagogical control and allow for non-destructive experimentation, the instructor manages the AI's output through a structured interactive workflow:

\textbf{Browsing Actionable Suggestions:} Upon processing the spoken pedagogical intent, the system populates a list of Interactive Action Cards on the LLM Interaction Panel (Figure \ref{fig:instructor interface}A2). Each card represents a discrete pedagogical step or a variation of the requested lighting setup, allowing the instructor to review the AI's interpretation before any scene changes occur.

\textbf{Apply and Revert Suggestions:} As shown in Figure \ref{fig:instructor interface}A2-A3, instructors can instantly execute or revert suggested lighting demos and annotations via one-click "Apply" and "Revert" functions. For high-efficiency coordination, "Apply All" and "Decline All" options are provided, ensuring the live demonstration remains fluid, responsive, and error-tolerant.

\textbf{Inspecting Detailed Parameters:} For advanced technical instruction, instructors can specify granular control parameters. For instance, an instructor can indicate a precise target position on the stage for a spotlight to track, bridging the gap between high-level intent and low-level execution.

\subsubsection{Intelligent Jargon Panel}
The Jargon Panel (Figure \ref{fig:instructor interface}D) serves as an ``on-demand'' conceptual bridge between professional terminology and student comprehension (\textbf{\textit{DR3}}).
\textbf{Transcription with Semantic Highlighting:} The instructional transcript is displayed with domain-specific keywords (e.g., ``Hue'' and ``Color wheel'') highlighted in a distinct color (Figure \ref{fig:instructor interface}D1).
\textbf{Context-Aware Definitions:} A concise list of definitions for the identified jargon is displayed in an adjacent panel (Figure \ref{fig:instructor interface}D2). This allows students to immediately associate technical terminology with the live demonstration, eliminating the need for the instructor to pause the lesson for basic definitions.
\\
\subsubsection{Manual Administrative Toolkit}
For global scene management and session control, the instructor is equipped with a persistent administrative toolkit (Figure \ref{fig:instructor interface}C):
\textbf{Pedagogical Support Tools:} To facilitate seamless instruction, \textit{LumiNote} provides three core utilities: (1) \textbf{Session Recording} (Figure \ref{fig:instructor interface}C1), which archives multimodal interactions for high-fidelity in-VR replay and review; (2) \textbf{Global Reset} (Figure \ref{fig:instructor interface}C2), enabling an instant return to default states for rapid transitions between modules; and (3) a \textbf{Laser Ray} (Figure \ref{fig:instructor interface}C3), a synchronized 3D pointer that serves as a ``virtual baton'' to direct student attention in real-time (\textbf{\textit{DR1}}).

\subsection{Technical Architecture}
\begin{figure*}[t]
    \centering
    \includegraphics[width=1\textwidth]{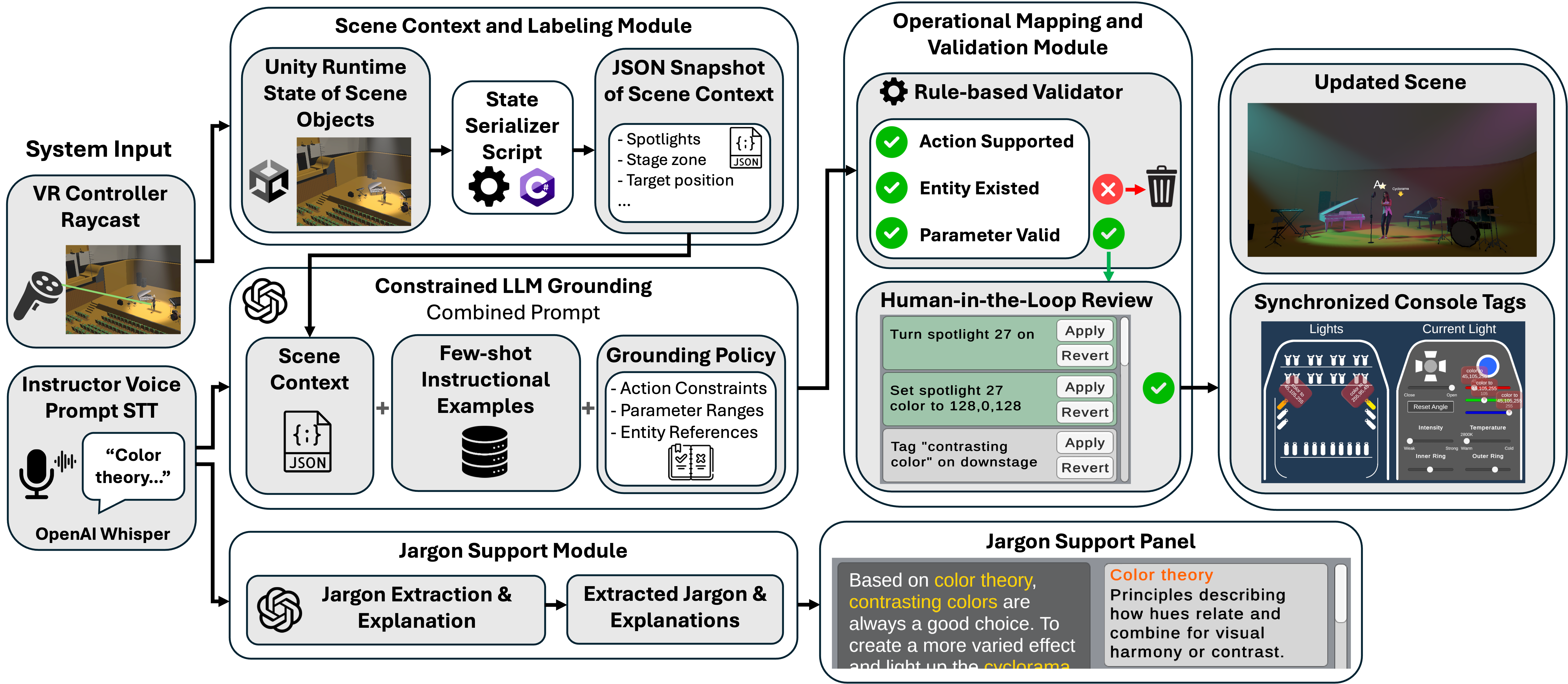}
    \caption{System Architecture}
    \label{fig:LumiNote architecture}
\end{figure*}

In this section, we present the architecture of LumiNote within our immersive instruction system (Figure \ref{fig:LumiNote architecture}). The architecture consists of four core components: the Scene Context and Labeling Module, Constrained LLM Grounding, Jargon Support Module, and Operational Mapping and Validation Module.

\subsubsection{Scene Context and Labeling Module}
The Scene Context and Labeling Module serves as the bridge between the immersive VR environment and the LLMs-based reasoning process by performing real-time state serialization. Prior work on 3D language grounding demonstrates that serializing physical scene graphs into structured textual snapshots allows LLMs to efficiently reason over spatial attributes with minimal processing latency \cite{rana2023sayplan}. Following this approach, since the Unity engine natively maintains the real-time state of all objects, this module focuses on extracting and filtering relevant Scene Metadata into a format optimized for LLM processing. The module serializes a lightweight JSON snapshot that captures the current properties of all lighting fixtures, including unique IDs, emission states, transform data (position/rotation), color, intensity, beam angles, and barn-door openness.

Beyond the lighting hardware, the module serializes the broader scene context. This includes the identifiers and position coordinates of performers and instruments, as well as zoned areas \cite{karen_different_2023} of the stage (e.g., center stage, upstage, downstage, stage left/right, apron, and cyclorama).

To resolve ambiguous or abstract spatial commands, the system employs an Iterative Laser Anchor mechanism. When the LLM detects a deictic reference in the user's verbal cue (e.g., ``Rotate the spotlight to here'') without a clear target position, the module triggers a system prompt that asks the user to specify the precise location. The user then utilizes the VR controller to point at the intended area. The system captures the resulting $Vector3$ intersection with the stage geometry and appends this coordinate to the metadata, ensuring the generative module understands both the pedagogical intent and the exact target within the 3D space.

\subsubsection{Constrained LLM Grounding}
The LLM grounding module translates instructor-provided lighting instructions into structured suggested actions that can be inspected and executed within the VR environment. The instructor's verbal input is first transcribed into text through the voice-driven interaction interface and serves as the high-level instructional intent for LLM reasoning. To support responsive interaction in VR, our prompt design uses direct few-shot in-context learning \cite{brown2020language} rather than computationally intensive multi-turn reasoning. The prompt combines three sources of grounding information: (1) \textit{runtime Scene Context}, including the current states of enabled fixtures (e.g., fixture IDs, colors, intensities, and beam parameters); (2) \textit{static Scene Metadata}, describing the theater layout, spatial relationships, fixture groups, and available objects; and (3) \textit{few-shot instructional examples} paired with a predefined action schema.

The few-shot examples are derived from existing stage lighting instructional materials and cover four foundational lighting concept categories \cite{collins_lookingatlight}: \textit{fixture/configuration}, which concerns manipulating individual fixture properties or states; \textit{spatial behavior}, which concerns relationships among fixtures, performers, and stage areas; \textit{visual effects}, which describe intended visual atmosphere or appearance; and \textit{composition strategies}, which concern multi-fixture coordination and overall stage composition. Rather than serving only as domain examples, these cases provide mappings between pedagogical intent and the system's executable action vocabulary. This allows the LLM to interpret domain-specific instructions while remaining within the operations supported by the VR environment, without requiring additional model training.

Given the instructor's intent, scene context, and action constraints, the LLM generates Structured Suggested Actions following the predefined schema. Each action specifies the operation type, target object, and required parameters, allowing an instructional intent to be grounded in the entities and operations available in the VR environment. For example, an instruction such as ``make the performer stand out with a warmer light'' can be translated into an executable color adjustment represented as \texttt{SET\_COLOR(Light\_27,[255,140,0])}.

\paragraph{Grounding Policy.}
Rather than allowing the LLM to directly control the virtual theater, LumiNote constrains generation along three dimensions: \textit{what can be changed}, \textit{what can be referenced}, and \textit{how changes are represented}. The predefined action schema specifies the supported operations and parameter ranges, while runtime scene context and static scene metadata constrain references to entities and spatial states that exist in the current environment. Few-shot examples further demonstrate how instructional intents can be expressed through these available operations. Together, these constraints bound the generative space to the operational vocabulary of the VR environment, reducing the possibility that the LLM proposes unsupported actions or invents capabilities that are not available in the system.

For transparency, the prompt constrains generation using a closed action vocabulary, bounded parameter ranges, scene-entity references, and few-shot mappings from pedagogical intents to executable action sequences. The complete prompt and representative examples are provided in Appendix~\ref{app:prompts_and_system_architecture}.

These constraints do not guaranty that the generated actions are technically or pedagogically appropriate. Instead, they establish a bounded proposal space in which generated actions can be subsequently validated and reviewed by the instructor before execution.

\subsubsection{Jargon Support Module}

In parallel with action generation, LumiNote employs a separate LLM-based jargon support pipeline to assist students in understanding professional lighting terminology. The module takes the transcribed instructor speech as input and uses a dedicated prompt that guides the LLM to identify domain-specific lighting terms and generate concise, context-aware explanations based on the ongoing instruction. The extracted terms and generated definitions are then synchronized with the transcript interface, where relevant jargon is highlighted and explanations are displayed alongside the transcript. This enables students to connect professional terminology with the instructor's real-time demonstration without interrupting the teaching flow.

\subsubsection{Operational Mapping and Validation Module}

The generated Suggested Actions are processed by the Operational Mapping and Validation Module before execution. The module verifies whether each action is executable by checking the supported action types, referenced scene entities (e.g., fixture IDs or stage objects), and parameter constraints (e.g., valid RGB or intensity ranges). Invalid or unsupported actions are filtered out before reaching the instructor.

The remaining actions are presented through a human-in-the-loop interface, allowing instructors to review, modify, apply, or reject suggestions before execution. After approval, the system updates the VR environment and generates synchronized instructional feedback. For lighting-related actions, LumiNote automatically creates console tags that annotate the corresponding virtual controls with the executed operation and parameter values, helping students connect visible lighting effects with underlying technical operations.

\subsection{Implementation}
We developed the prototype VR system 
built using the Unity3D platform (version 6000.0.23f1) and is deployed on the Meta Quest 3 standalone headset. To facilitate natural interaction, the user's voice input is transcribed into text using the OpenAI Whisper STT model. For the core logic and instruction generation, we utilized the GPT-4o model to process the transcribed text and scene context. The system backend is powered by the Django framework, with a SQLite database used to record interaction logs and in-scene lighting configurations.

\section{Exploratory Study}
We conducted a two-phase exploratory study to understand when and how stage lighting instructors incorporate LLM assistance into live VR teaching and how learners receive the resulting instructional content. Since LumiNote is primarily instructor-facing and augments pedagogical delivery rather than mediating dynamic student interaction, we adopted this two-phase design. Phase 1 examined how instructors incorporated LLM assistance into familiar teaching topics, using a \textit{No LLM} session as a reference for each instructor's manual teaching flow. Phase 2 used recorded VR teaching sessions to examine learner reception of the resulting instructional content while controlling for instructor identity and delivery.

The study focused on three research questions: (1) when is LLM assistance most useful during live VR instruction, and what grounding challenges arise across different types of instructional intent? (2) how do instructors incorporate, control, and refine LLM-generated support within their teaching flow, and how does this assistance reshape their instructional work? and (3) how do learners receive the resulting demonstrations and representations, and how do their preferences align with those of instructors?



\subsection{Participants}
For Phase 1, we recruited three stage lighting instructors (2M, 1F) from the technical theater arts industry and universities (T1–T3): a theater arts practitioner with four years of industry mentoring experience, a guest lecturer with over five years of professional practice and one year of college teaching experience in lighting design, and a full-time college lecturer with three years of teaching experience in technical theater arts. The participants had a diverse range of VR familiarity: one instructor identified as a beginner, one as intermediate, and one as advanced. This variation provided cases with different levels of prior VR familiarity. Stage lighting education is a specialized professional domain, so instructor participants were required to have direct teaching experience and to be able to assess whether generated demonstrations aligned with authentic instructional practice. Following prior HCI work emphasizing the importance of domain expertise in evaluating systems for complex professional settings~\cite{chilana2010understanding,CHI22CineData}, we prioritized close domain fit and in-depth exploratory observation in recruiting the instructor sample.

For Phase 2, we recruited 24 students (10M, 14F) from the university campus and via social media, with ages ranging from 18 to 31 ($M=22.62, SD=4.15$). The students had diverse backgrounds, ranging from bachelor's to doctoral degrees across various fields of study. All student participants reported having ``none'' or ``beginner'' levels of VR experience. Additionally, over 80\% of the students reported being ``not familiar at all'' or ``slightly familiar'' with the stage lighting concepts covered in the tutorial.

This study was approved by the Institutional Review Board. 
All participants provided informed consent prior to participation.

\subsection{Procedures}

\subsubsection{Phase 1: Instructor Teaching Sessions}
Each instructor completed two VR teaching sessions: a session without LLM assistance (\textit{No LLM}) followed by a session with full LumiNote support (\textit{With LLM}). The \textit{No LLM} session served as a reference for each instructor's manual teaching flow, while the subsequent \textit{With LLM} session allowed us to observe where and how LLM assistance was incorporated after the instructor had taught the same topics once without generative support. Before the study, instructors received a 15--20 minute onboarding session to familiarize themselves with LumiNote's core interactions and LLM-assisted functions.

In both conditions, instructors taught the same four predefined beginner-level lighting concepts: side lighting, color dimensions, color combinations, and emotional atmosphere. These topics were selected to cover the foundational learning dimensions, including fixture/configuration, spatial behavior, visual effects, and composition strategies, while remaining suitable for a short beginner-level teaching session. The detailed teaching protocol and concept scripts are provided in Appendix~\ref{app:instructor_teaching_protocal}. In the \textit{No LLM} condition, the LLM module was disabled. In the \textit{With LLM} condition, instructors were encouraged to use the system to generate spatial annotations, request on-demand lighting demonstrations, and obtain linguistic support for explaining domain-specific terminology.

All teaching sessions were recorded, and system interaction logs were collected. After each session, instructors completed post-session questionnaires. After completing both sessions, instructors participated in a semi-structured interview focusing on communication burden, session preference, feature utility, VR adoption willingness, teaching prompts and reminders, and future interaction patterns.


\subsubsection{Phase 2: Learner Reception and Immediate Task Performance}
Students participated individually and were randomly assigned to two conditions ($n=12$ each): \textit{No LLM Recording} or \textit{With LLM Recording}. Both recordings were created by the same instructor, selected from the three Phase 1 instructors, so that instructor identity and teaching style were held constant across conditions. The paired recordings were selected based on predefined criteria, including alignment with the target learning objectives, acceptable audio--visual quality, and comparable duration. Phase 2 focused on learner reception and immediate task performance rather than spontaneous instructor--student interaction. Students experienced the recorded demonstrations after the instructor had reviewed and applied any generated actions, and students did not directly trigger or modify LLM outputs. Thus, this phase examined learner-facing effects of the instructional content created by the instructor while controlling for instructor identity and delivery; it does not evaluate spontaneous live classroom interaction.

Students then completed four sequential practical lighting tasks designed to assess whether they could apply concepts demonstrated in the lesson. The first three tasks focused on specific taught concepts: adding a directional sidelight to make the solo performer stand out from the background, adjusting light quality by increasing edge-transition hardness and shadow sharpness, and applying warm–cool color contrast across active fixtures. The final task was an open-ended composition task in which students designed a complete lighting state for a solo singer at center stage to create a lyrical performance atmosphere. This task required students to integrate rhythm, color harmony, spatial composition, and emotional expression rather than reproduce a single demonstrated operation.

After completing the final scene, students provided a verbal rationale explaining their design choices and the lighting concepts they attempted to apply. Each submission therefore consisted of a high-resolution screenshot of the completed VR lighting scene and the student's rationale. Expert instructors later evaluated these submissions using a rubric assessing concept accuracy, visual atmosphere, and concept comprehensiveness. Students then completed post-session questionnaires and a brief semi-structured interview.

All student interactions (lighting parameter adjustments, object selections, and tool usage), task completion time, and task responses were logged.

\subsection{Data Sources and Analytic Role}

For the instructor phase, post-study interviews, session recordings, and observed teaching behaviors provided the primary interpretive evidence for understanding when and how instructors incorporated LLM assistance into their teaching. System logs captured instructor prompts, generated suggestions, instructor decisions, and follow-up actions during the \textit{With LLM} sessions.

Questionnaire and behavioral measures were used as supporting evidence. For instructors, we collected SUS\cite{brooke1996sus}, NASA-TLX\cite{hart2006nasa}, custom ratings of immediacy, clarity, instructional effectiveness, professionalism, pedagogical capability, feature-level ratings, and session duration. Given the small instructor sample, these measures were interpreted descriptively to contextualize reported changes in workload, teaching flow, and perceived instructional support rather than as population-level evidence.

For the learner phase, student interviews and open-ended responses provided the primary interpretive evidence for learner reception. We interpreted these responses alongside feature ratings, SUS, NASA-TLX, IPQ presence\cite{schubert2001experience}, learning-related self-ratings, task completion time, expert task scores, and verbal rationales to examine how students received the instructional content created by the instructor and how their responses compared with instructors' perceptions of the same representations.

\subsection{Analysis}
\subsubsection{Qualitative Analysis}

We reviewed instructor interviews, session recordings, observed teaching behaviors, and open-ended responses to identify recurring patterns relevant to the first two research questions. The analysis focused on when instructors invoked LLM support, how they responded to generated suggestions, how they maintained control over the teaching flow, and how they described changes in operational effort and pedagogical expression.

Student interviews and open-ended responses were analyzed separately to characterize learner reception. We examined which instructional cues students found useful, what remained difficult to follow, and how the recorded instruction supported their subsequent tasks. We then compared instructor- and learner-side observations where relevant to identify similarities and differences in the perceived usefulness of instructional representations.

\subsubsection{Instructor Prompt Categorization and Suggested-Action Analysis}
To characterize how different forms of instructional intent were translated into executable lighting actions, we conducted a two-level analysis of the LLM-generated suggestions collected during the instructor sessions.

First, at the prompt level, we categorized each instructor prompt according to its dominant instructional intent: fixture/configuration, visual effects, spatial behavior, composition strategies, or ambiguous. Prompts were labeled as ambiguous when their intended target, parameter, or desired outcome could not be determined reliably from the recorded transcription.

Second, at the suggested-action level, we examined the LLM-generated actions associated with each instructor prompt. Because a single prompt could produce multiple suggested actions, we analyzed prompt--suggested action pairs and recorded whether each suggestion was applied, rejected, or modified by the instructor. We interpret action-level adoption as an interaction-level indicator of whether a generated suggestion was incorporated into the ongoing teaching flow, rather than as a measure of model accuracy or pedagogical correctness. For rejected or modified suggestions, we further examined the source of misalignment and instructors' subsequent actions.

\subsubsection{Quantitative Analysis}

Quantitative measures were analyzed primarily as supporting evidence for the exploratory findings. For the instructor phase, given the small sample ($n=3$) and the fixed \textit{No LLM} to \textit{With LLM} sequence, questionnaire and behavioral measures were summarized descriptively rather than subjected to inferential testing. We report means and standard deviations for NASA-TLX, SUS, custom instructional ratings, feature-level ratings, and session duration, and use these measures to contextualize patterns identified from instructor interviews and observed teaching behaviors.

For the learner phase, we conducted between-subject comparisons between the \textit{No LLM Recording} and \textit{With LLM Recording} conditions for questionnaire and task measures, including SUS, NASA-TLX, IPQ presence, learning-related self-ratings, task completion time, and expert task scores. Independent-samples $t$-tests were used for between-condition comparisons, with Cohen's $d$ reported as an effect-size measure. Feature-level usefulness ratings, which were collected only in the \textit{With LLM} condition, were summarized descriptively using means and standard deviations.

\section{Results}
\label{sec:results}



Across the three \textit{With LLM} instructor sessions, instructors issued 55 unique prompts, which produced 531 prompt--suggested action pairs. The amount of LLM interaction varied substantially across instructors: T1 issued 26 prompts associated with 397 generated actions, T2 issued 21 prompts associated with 67 actions, and T3 issued 8 prompts associated with 67 actions. Across all 531 action pairs, 380 actions (71.6\%) were applied, 90 (16.9\%) were rejected, and 61 (11.5\%) were modified.

Because both the number and type of generated actions differed across instructors, we report the interaction-log results at two complementary levels. Table~\ref{tab:instruction_action_alignment} summarizes the pooled distribution and adoption patterns across instructional categories, while Table~\ref{tab:individual_llm_adoption} shows how these patterns varied across T1--T3.

Visual effects formed the largest prompt category (27 of 55 prompts, 49.1\%) and showed relatively high direct action-level adoption (212 of 245 actions, 86.5\%). Composition strategies also showed relatively high pooled adoption (69 of 89 actions, 77.5\%), although adoption varied more across instructors. Fixture/configuration suggestions showed the lowest pooled direct adoption among the four non-ambiguous categories (38 of 66 actions, 57.6\%), but this pattern was strongly instructor-dependent: T1 applied 27 of 28 fixture/configuration actions, whereas T3 applied 10 of 37. Spatial-behavior suggestions showed similarly uneven instructor-level distributions.

Six prompts were categorized as ambiguous because their intended target, parameter, or desired outcome could not be determined reliably from the recorded transcription. These prompts generated 92 action pairs and were analyzed separately from the four interpretable lighting-content categories. Among rejected and modified actions, the most common issues involved unintended repeated interactions, directional-reference interpretation, speech-transcription problems, and instructions that instructors reformulated or no longer considered necessary. These cases indicate that action status should be interpreted as whether a suggestion entered the ongoing teaching flow rather than as a direct measure of model accuracy.

\begin{table*}[t]
\centering
\small
\caption{Prompt-level distribution and action-level adoption of LLM-generated suggestions. Prompt outcomes are reported as AA/AR/PA/M, denoting all accepted, all rejected, partially accepted, and modified/other.}
\label{tab:instruction_action_alignment}
\begin{tabular}{lcccccc}
\toprule
\textbf{Category} &
\textbf{Prompt n (\%)} &
\textbf{Prompt outcome} &
\textbf{Action pairs} &
\textbf{Applied} &
\textbf{Rejected} &
\textbf{Modified/Other} \\
\midrule

Visual effects
& 27 (49.1\%)
& 25/1/1/0
& 245
& 212 (86.5\%)
& 33 (13.5\%)
& 0 (0.0\%) \\

Composition strategies
& 11 (20.0\%)
& 8/2/0/1
& 89
& 69 (77.5\%)
& 13 (14.6\%)
& 7 (7.9\%) \\

Fixture/configuration
& 7 (12.7\%)
& 4/0/3/0
& 66
& 38 (57.6\%)
& 28 (42.4\%)
& 0 (0.0\%) \\

Spatial behavior
& 4 (7.3\%)
& 2/0/2/0
& 39
& 29 (74.4\%)
& 9 (23.1\%)
& 1 (2.6\%) \\

Ambiguous
& 6 (10.9\%)
& 2/2/0/2
& 92
& \multicolumn{3}{c}{Analyzed separately as ambiguity-resolution cases} \\

\midrule

Non-ambiguous total
& 49 (89.1\%)
& 39/3/6/1
& 439
& 348 (79.3\%)
& 83 (18.9\%)
& 8 (1.8\%) \\

All prompts
& 55 (100.0\%)
& 41/5/6/3
& 531
& 380 (71.6\%)
& 90 (16.9\%)
& 61(11.5\%) \\

\bottomrule
\end{tabular}
\end{table*}

\begin{table*}[t]
\centering
\small
\caption{Instructor-level distribution and adoption of LLM-generated suggestions across instructional categories. Percentages for Applied, Rejected, and Modified are calculated within each instructor--category combination. ``--'' indicates that the instructor did not issue prompts in that category.}
\label{tab:individual_llm_adoption}
\begin{tabular}{llrrrrr}
\toprule
\textbf{Instructor} &
\textbf{Category} &
\textbf{Prompt n} &
\textbf{Action n} &
\textbf{Applied} &
\textbf{Rejected} &
\textbf{Modified} \\
\midrule

T1 &
Visual effects &
12 & 208 &
175 (84.1\%) &
33 (15.9\%) &
0 (0.0\%) \\

&
Composition strategies &
2 & 43 &
37 (86.0\%) &
6 (14.0\%) &
0 (0.0\%) \\

&
Fixture/configuration &
3 & 28 &
27 (96.4\%) &
1 (3.6\%) &
0 (0.0\%) \\

&
Spatial behavior &
3 & 26 &
16 (61.5\%) &
9 (34.6\%) &
1 (3.8\%) \\

&
Ambiguous &
6 & 92 &
32 (34.8\%) &
7 (7.6\%) &
53 (57.6\%) \\

\cmidrule(lr){2-7}
&
\textit{T1 Total} &
26 & 397 &
287 (72.3\%) &
56 (14.1\%) &
54 (13.6\%) \\

\midrule

T2 &
Visual effects &
13 & 27 &
27 (100.0\%) &
0 (0.0\%) &
0 (0.0\%) \\

&
Composition strategies &
7 & 39 &
25 (64.1\%) &
7 (17.9\%) &
7 (17.9\%) \\

&
Fixture/configuration &
1 & 1 &
1 (100.0\%) &
0 (0.0\%) &
0 (0.0\%) \\

&
Spatial behavior &
-- & -- & -- & -- & -- \\

&
Ambiguous &
-- & -- & -- & -- & -- \\

\cmidrule(lr){2-7}
&
\textit{T2 Total} &
21 & 67 &
53 (79.1\%) &
7 (10.4\%) &
7 (10.4\%) \\

\midrule

T3 &
Visual effects &
2 & 10 &
10 (100.0\%) &
0 (0.0\%) &
0 (0.0\%) \\

&
Composition strategies &
2 & 7 &
7 (100.0\%) &
0 (0.0\%) &
0 (0.0\%) \\

&
Fixture/configuration &
3 & 37 &
10 (27.0\%) &
27 (73.0\%) &
0 (0.0\%) \\

&
Spatial behavior &
1 & 13 &
13 (100.0\%) &
0 (0.0\%) &
0 (0.0\%) \\

&
Ambiguous &
-- & -- & -- & -- & -- \\

\cmidrule(lr){2-7}
&
\textit{T3 Total} &
8 & 67 &
40 (59.7\%) &
27 (40.3\%) &
0 (0.0\%) \\








\bottomrule
\end{tabular}
\end{table*}

\subsection{When LLM Assistance Was Most Useful and Where Grounding Became Difficult}

\subsubsection{Expressive and Under-Specified Intent Provided a Particularly Useful Role for LLM Assistance}

A recurring pattern in instructor interviews was that LLM assistance was particularly valuable when the desired lighting effect was expressive, under-specified, or difficult to define through precise parameters in advance. T2 described LumiNote as helpful ``when I wanted to create a lighting effect that was not very explicit and was more feeling-based,'' and estimated that the generated result was ``about 70\% aligned'' with the intended effect. T2 further noted that prompting made it possible to discuss and try more ambiguous or creative ideas beyond a basic syllabus. Similarly, T1 noted that LumiNote could quickly generate a complete lighting effect, reducing the time needed during teaching, while T3 preferred the \textit{With LLM} session because the intended visual effect could be presented quickly.

The interaction logs provided complementary evidence for this pattern. Visual effects and composition strategies together accounted for 38 of the 55 instructor prompts and showed relatively high pooled action-level adoption. In contrast, fixture/configuration prompts were less frequent and showed lower pooled adoption. These observations suggest that instructors drew most strongly on LLM assistance when teaching perceptual, expressive, or composition-oriented ideas for which rapid visual externalization was useful.

\subsubsection{Precise Fixture-Level Requests Exposed Greater Grounding Demands}

More operationally precise requests exposed a different challenge. Fixture/configuration suggestions showed the lowest pooled direct adoption rate among the non-ambiguous categories, although the pattern varied substantially across instructors. A closer review of the 28 rejected fixture/configuration actions showed that 26 were associated with directional-reference interpretation. In these cases, the LLM often selected the wrong fixture when interpreting phrases such as ``left light'' or ``side light.''

These cases suggest that fixture-level grounding depended not only on whether an action was technically executable, but also on alignment with the current scene state, fixture identity, spatial reference frame, and the instructor's intended configuration. Thus, as instructional intent became more operationally specific, successful assistance required more precise grounding in the state and spatial organization of the scene.

\subsection{How Instructors Integrated and Controlled LLM Assistance}

\subsubsection{LLM Assistance Functioned as a Controllable Refinement Process}

Instructors did not treat generated suggestions as finished teaching plans. Instead, they incorporated LumiNote within their own teaching goals and subgoals. T3 emphasized that the lesson still needed to follow a planned teaching flow with an overall goal divided into subgoals. When generated outputs did not match the intended meaning, T3 would skip them or prompt again to obtain a configuration closer to the intended explanation. This reflects a control-oriented use pattern in which generation served the instructor's pedagogical plan rather than determining it.

Follow-up behavior further illustrates this refinement process. Among 147 rejected or modified suggestions with recorded follow-up actions, 127 (86.4\%) were followed by a new prompt, 19 (12.9\%) with no additional action, and only 1 (0.7\%) involved direct manual adjustment. Reprompting was commonly associated with insufficiently specific instructions, spatial-reference problems, or speech-transcription issues. Overall, instructor--LLM interaction unfolded as an iterative process in which instructors progressively clarified their intent while retaining authority over the resulting configuration.

Instructor feedback also pointed toward more structured forms of control. All three instructors expressed interest in templates or pre-set lighting combinations tied to particular concepts, suggesting a desire to reduce repetitive prompting while preserving the ability to inspect and select generated outcomes.

Use patterns differed substantially across instructors. T1, with one year of teaching experience, triggered substantially more generated actions than T2 and T3, who had four and three years of teaching experience, respectively. T2 suggested that LumiNote might be especially useful for instructors who are still developing familiarity with their teaching materials, while T3 described experienced instructors as more likely to follow an established internal teaching plan. Given the small instructor sample, these observations suggest a possible relationship between instructor experience and generative-tool use rather than a systematic expertise effect.

\subsubsection{LLM Assistance Shifted Effort from Manual Setup to Pedagogical Expression}

Instructors consistently described LumiNote as reducing the effort required to prepare, adjust, and operate lighting demonstrations during instruction. T2 preferred the \textit{With LLM} session because it reduced the need to manually prepare and tune lighting demos, allowing the instructor to focus on the intended effect rather than operational details. T1 similarly noted that LumiNote saved time in preparing, selecting, and operating demonstrations. T3 described the system as allowing the instructor to state a desired outcome without manually realizing each lighting combination.

Questionnaire and behavioral measures contextualized this perceived shift. Instructor NASA-TLX scores decreased descriptively from 3.50 ($SD=1.88$) to 2.22 ($SD=1.11$), and mean session duration decreased from 19 minutes and 34 seconds to 12 minutes and 33 seconds. Instructor SUS increased from 65.83 ($SD=16.27$) to 79.17 ($SD=7.64$). Ratings also increased descriptively for immediacy (\textit{No LLM}: 4.83, \textit{With LLM}: 5.83), clarity (\textit{No LLM}: 4.67, \textit{With LLM}: 6.00), instructional effectiveness (\textit{No LLM}: 4.50, \textit{With LLM}: 5.83), professionalism (\textit{No LLM}: 5.00, \textit{With LLM}: 5.50), and pedagogical capability (\textit{No LLM}: 5.17, \textit{With LLM}: 6.00). Given the small instructor sample and fixed condition order, these measures are interpreted as contextual evidence rather than inferential evidence of improvement. Together with instructor accounts, they are consistent with a shift from operational coordination toward attention to what instructors wanted to express and how they wanted to present it.

\subsection{Learner Reception and Representation Alignment}

\subsubsection{Instructor and Learner Preferences Revealed Representation-Alignment Tensions}

The learner phase allowed us to examine whether representations that supported instructor explanation were also useful for novice learners. Instructors rated AI-generated lighting demonstrations and directional arrows as the most effective functions for instruction, both averaging 6.0 on a 7-point scale. T1 and T3 emphasized that directional arrows helped externalize lighting relationships that can be difficult to communicate verbally, such as light direction and beam behavior. In contrast, instructors rated console action tags and the animated instructor avatar relatively lower, both averaging 5.0.

Students showed a different pattern. In the \textit{With LLM} condition, students rated the green laser pointer as the most helpful feature ($M=5.50$, $SD=1.17$), followed by console panel tagging with suggested actions and values ($M=5.33$, $SD=1.15$) and lighting demonstrations ($M=5.25$, $SD=0.87$). Spatial arrow annotations and the animated instructor avatar were rated more moderately ($M=4.58$, $SD=1.24$; $M=4.58$, $SD=1.31$), while star-shaped highlights received the lowest ratings ($M=4.33$, $SD=1.30$).

This contrast suggests that the same instructional representation can serve different roles for instructors and learners. For instructors, directional arrows helped externalize expert spatial reasoning. For novice learners, the most highly rated cues instead directed attention toward a concrete location (e.g., S6: "Theater is large, it's very hard to intuitively understand which part to focus on, so laser can help") or connected a visible lighting effect to the corresponding console operation (e.g., S12: "The lighting demonstration with tagging on the buttons and sliders are helpful for me to understand what to do exactly."). Learner-facing representations therefore need to do more than reproduce expert expression; they also need to help learners identify where to look, what changed, and which operation produced that change.

\begin{figure*}
    \centering
    \includegraphics[width=1\textwidth]{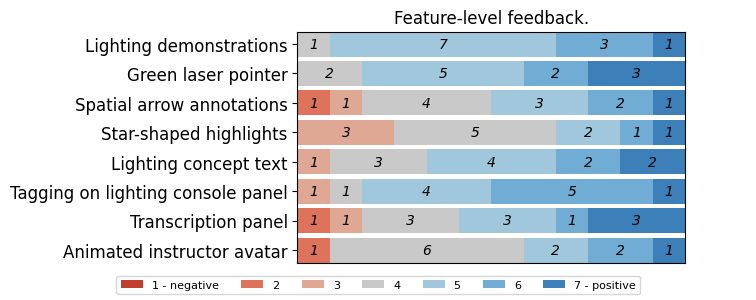}
    \caption{Students in the \textit{With LLM} condition rated the usefulness of individual LumiNote features on a 7-point Likert scale.}
    \label{fig:student_usefulness}
\end{figure*}

\subsubsection{Learner Outcomes Reflected Presence and Technical Articulation Rather Than Immediate Performance Gains}

We further examined how the resulting instructional content was reflected in students' broader experiences and immediate task outcomes. Students reported higher presence in the \textit{With LLM} condition ($M=5.72$, $SD=0.82$) than in the \textit{No LLM} condition ($M=4.93$, $SD=0.80$; $t(22)=-2.40$, $p=.025$, $d=0.98$). SUS, NASA-TLX, subjective learning ratings, task completion time, and expert task scores showed no significant between-condition differences.

Student rationales nevertheless differed qualitatively. Eight of the twelve students in the \textit{With LLM} condition described specific lighting techniques, such as ``blurring the inner ring to reduce hardness'' or using contrasting colors for better harmony, whereas nine of the twelve students in the \textit{No LLM} condition relied more on abstract or intuitive descriptions, such as ``personal feeling.'' These observations suggest that the learner-facing instructional cues may have influenced students' sense of presence and the technical language they used to articulate their lighting decisions, even though immediate task performance remained comparable.

\begin{figure*}
    \centering
    \includegraphics[width=\textwidth]{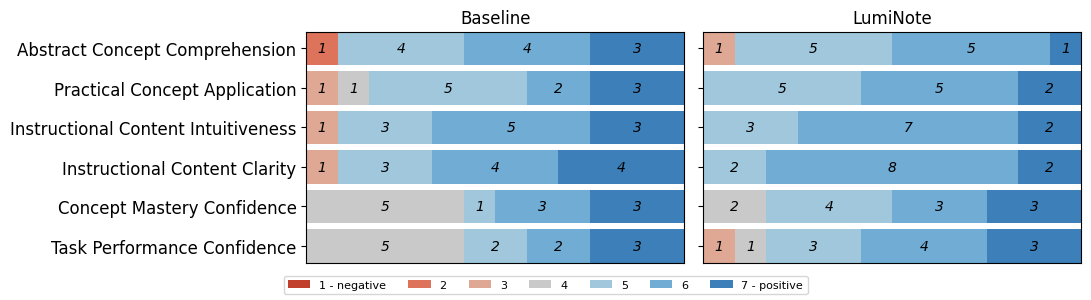}
    \caption{Students provided post-session ratings on six learning-related items using a 7-point Likert scale.}
    \label{fig:student_learning}
\end{figure*}

\section{Discussion}


\subsection{Generative Assistance Is Most Useful at the Intent-to-Operation Gap}

Our findings suggest that generative assistance is not equally valuable across instructional intents. LumiNote was most useful when instructors could articulate the perceptual or expressive outcome they wanted to demonstrate, but had not fully specified the fixture-level operations required to produce it. In these situations, the LLM could propose an operational realization of an expert's high-level intent, allowing instructors to externalize ideas such as atmosphere, color relationships, or visual composition without manually specifying every parameter.

By contrast, when instructors already had a precise operational target in mind, the challenge shifted from generation to grounding. Fixture-level requests depended on correctly identifying scene entities, interpreting spatial references, and aligning generated actions with the instructor's intended configuration. This distinction suggests that the value of generative assistance lies less in replacing well-specified operations than in bridging an \textit{intent-to-operation gap}, where experts know what they want to communicate but have not yet articulated every action needed to realize it.

\textbf{Design Implication: Adaptive Support Based on Intent Specificity.}
Future instructor-facing systems could vary the form of AI assistance according to the specificity of the instructor's intent. For expressive or under-specified requests, the system could propose alternative realizations that instructors can quickly inspect and compare. For precise operational requests, the system could instead prioritize reference disambiguation, spatial anchoring, explicit parameter confirmation, or deterministic controls. Such an approach would treat generation and grounding as complementary forms of support rather than assuming that open-ended generation is equally appropriate for every instructional task.

\subsection{LLM Assistance as Controllable Refinement Rather Than Autonomous Instruction}

The instructor interactions in our study suggest that LLM-generated content was most useful when treated as a proposal within an existing pedagogical plan rather than as an autonomous teaching decision. Instructors retained their own goals and subgoals, selectively applied generated suggestions, and commonly reformulated prompts when an output did not match the intended explanation. The LLM therefore participated in an iterative refinement process in which instructors progressively externalized and clarified their intent while retaining authority over what entered the lesson.

This interaction pattern also helps explain the observed shift from manual operation toward pedagogical expression. When the system handled some of the work required to instantiate demonstrations, instructors could devote more attention to the effect or concept they wanted to communicate. Importantly, this redistribution of effort did not remove instructor decision making. Instead, it moved instructor effort away from low-level configuration toward selecting, evaluating, and refining candidate demonstrations.

\textbf{Design Implication: Structured Interaction for Controllable Generation.}
Instructor feedback suggests that controllable refinement may benefit from more structured interaction than unrestricted conversational prompting alone. Templates, reusable lighting combinations, or other modular forms of input could allow instructors to express recurring instructional intents without repeatedly reformulating similar prompts. Likewise, generated outputs could be organized as discrete, inspectable modules that can be selected, modified, or rejected. This motivates exploring modular and declarative interaction for domain-specific LLM support~\cite{declare}. Rather than positioning the LLM as an autonomous agent, such interfaces can treat generation as a bounded proposal mechanism whose outputs remain subject to expert judgment. This distinction is particularly important in professional instruction, where technical executability does not necessarily imply pedagogical appropriateness.

\subsection{Mediating Between Expert Expression and Learner Comprehension}

Our findings also highlight a second mediation problem between how experts express instructional knowledge and how novices receive it. Instructors valued representations such as directional arrows because they helped externalize spatial relationships that are implicit in expert reasoning. Learners, however, more highly valued cues such as the laser pointer and console tagging that directed attention to a concrete location or connected a visible effect to the operation that produced it. The representation that best supports expert explanation therefore may not be the representation that best supports novice interpretation.

This distinction suggests that learner-facing representation should not be treated as a direct copy of expert-facing expression. Instead, the system may need to preserve the instructor's intended concept while translating how that concept is visually and technically presented to learners. For example, an instructor-facing representation may emphasize a spatial relationship across several fixtures, whereas a learner-facing representation may need to emphasize where to look, what changed, and which control produced the change.

The learner-side outcomes further reinforce the need for a cautious interpretation of instructional representation. The \textit{With LLM} condition was associated with higher presence and more technically specific rationales, while immediate task performance remained comparable across conditions. This pattern suggests that richer instructional representations may first influence how learners experience and articulate a concept without necessarily producing immediate performance gains. Longer-term studies are needed to determine whether such differences contribute to retention or transfer.

\textbf{Design Implication: Instructor--Learner Representation Alignment.}
Future immersive instructional systems could separate the representation used to externalize expert reasoning from the representation ultimately presented to learners. Learner-facing representations could be adapted according to the current instructional goal, the learner's expertise, and whether the task requires spatial understanding, attention guidance, or connection between an observed effect and its underlying operation. In this view, the LLM is not only an intent-to-action translator, but potentially a mediation layer between expert expression and learner comprehension.

\subsection{Transfer Conditions for Expert-in-the-Loop Grounding}

Building on these findings, we identify four conditions that characterize when this interaction pattern may be applicable beyond stage lighting. These conditions concern the relationship among expert intent, executable environment state, bounded generation, and human oversight rather than the specific domain itself.

\textbf{Condition 1: Expert intent is under-specified.}
The interaction pattern is most relevant when experts can articulate what they want learners to understand or observe without specifying all of the operational steps needed to realize it. This creates an intent-to-operation gap that a generative system can help bridge. In stage lighting, for example, an instructor may express an intended atmosphere or visual effect without specifying the exact fixture and parameter changes in advance. The value of generation therefore lies in proposing an operational interpretation of expert intent rather than replacing an already well-specified procedure.

\textbf{Condition 2: The environment contains executable state.}
The environment must expose structured state that can be referenced and manipulated through system-supported operations. In LumiNote, this includes identifiable fixtures, stage objects, spatial relationships, and lighting parameters. Such state allows natural-language intent to be grounded in the current environment rather than interpreted as unconstrained text. Without executable environmental state, a system could generate descriptions or suggestions but could not reliably connect expert intent to situated actions.

\textbf{Condition 3: The domain has a bounded operational vocabulary.}
Generated actions must be constrained to a finite set of supported operations and valid parameter ranges. LumiNote uses a predefined action schema and scene metadata to restrict what the LLM can propose. This boundedness does not guarantee technical or pedagogical correctness. Instead, it limits generation to actions that can be inspected, validated, and potentially executed within the environment. Domains with similarly structured operational vocabularies may therefore provide suitable settings for this interaction pattern.

\textbf{Condition 4: Experts retain authority over generated actions.}
Generated actions should remain proposals rather than autonomous decisions. Experts must be able to review, modify, apply, reject, or regenerate suggestions according to their instructional goals. Grounding is therefore not complete when an action is merely executable; the expert must also determine whether the action preserves the intended pedagogical meaning.

Taken together, these conditions define a transferable interaction pattern for LLM-assisted immersive instruction: under-specified expert intent is translated into bounded, executable proposals that are grounded in the current environment, reviewed by the expert, and expressed through learner-facing representations. For example, in immersive chemistry instruction, an instructor's intent to demonstrate a titration endpoint could be grounded in the available laboratory equipment and translated into reviewable actions such as highlighting relevant instruments or visualizing an expected observation. The broader implication is not that LLMs can autonomously conduct instruction, but that immersive environments with structured executable state and bounded operations can provide a controllable substrate for mediating between expert intent and learner-facing representation.

\subsection{Limitations and Future Work}

Several limitations should be considered. First, our evaluation involved a small sample of instructors ($n=3$) and students ($n=24$), with a relatively narrow range of instructor experience and mostly novice learners. The findings therefore provide an exploratory account rather than population-level evidence. Second, the instructor phase followed a fixed \textit{No LLM}-to-\textit{With LLM} order, so descriptive differences in workload, session duration, and ratings may partly reflect practice or order effects. Third, the learner phase used recordings created by a single instructor to control delivery across conditions, and the \textit{With LLM} condition bundled multiple generated demonstrations and instructional representations. Learner-side findings may therefore depend on instructor-specific delivery and do not isolate the contribution of individual features or the LLM itself. Finally, our findings are tied to a particular technical implementation, and instructor acceptance, rejection, and modification were treated as interaction-level indicators of alignment rather than objective measures of technical or pedagogical correctness.

Future work should examine LumiNote across broader instructor and learner populations, multiple instructors, and longitudinal live-teaching settings. Controlled ablations could help isolate the contribution of different generated representations, while alternative grounding strategies, structured interaction mechanisms, and adaptive learner-facing representations could further clarify when and how LLM assistance supports immersive professional instruction.

\section{Conclusion}

We presented LumiNote, an instructor-facing LLM-assisted VR system for Stage Lighting Education that translates spoken pedagogical intent into reviewable scene actions and learner-facing representations. Through a two-phase exploratory study with 3 instructors and 24 students, we found that LLM assistance was particularly useful for expressive and under-specified lighting intent, whereas precise fixture-level requests required greater grounding and expert intervention. Instructors used generated suggestions as part of a controllable refinement process that shifted effort from manual demonstration setup toward pedagogical expression. Learner-side findings further revealed a representation-alignment tension between cues that supported expert expression and those novices found easiest to follow. Together, these findings characterize LLM-assisted immersive instruction as a domain-grounded mediation process among expert intent, executable scene operations, instructor control, and learner-facing representation.
\bibliographystyle{ACM-Reference-Format}
\bibliography{sample-base}

\appendix
\section{Appendix}

\subsection{Formative Study Codebook}
\label{app:formative_codebook}

Table~\ref{tab:formative_codebook} summarizes the final codebook used in the formative study analysis. The codebook includes current-practice codes, pedagogical-workflow codes, VR-affordance codes, desired-feature codes, and challenge codes. We used these codes to derive the design requirements reported in Section~3.3.
\begin{table*}[h]
\centering
\small
\caption{Condensed formative study codebook.}
\label{tab:formative_codebook}
\renewcommand{\arraystretch}{1.15}
\begin{tabular}{p{0.18\textwidth}p{0.30\textwidth}p{0.34\textwidth}p{0.10\textwidth}}
\toprule
\textbf{Code group} & \textbf{Representative codes} & \textbf{Operational definition} & \textbf{Design link} \\
\midrule

Current feedback practice
& Face-to-face discussion; visualization-supported discussion; real-time lab/artistry feedback; LMS verbal/written feedback.
& How instructors currently provide feedback to students across in-person, digital, and hands-on teaching settings.
& Scope \\
\midrule

Current tools
& Pre-visualization platform; LMS; audio recorder.
& Tools or platforms currently used for lighting design instruction, feedback, documentation, or post-lesson review.
& Scope \\
\midrule

Teaching workflow
& Light rigging, focusing, and plotting; basic concepts first, then production-driven learning; theory, practical, and artistry; lecture, practice, and production.
& Typical progression of stage lighting instruction from foundational theory to practical operation and artistic/production application.
& Scope \\
\midrule

VR affordances
& Showing equipment; safe for novice learners; real space to try; no need to accommodate performers; realistic 3D experience; remote collaboration; venue understanding; fast-paced workflow.
& Perceived advantages of VR for stage lighting education, especially for simulating venues, equipment, performers, and repeated practice.
& DR1, DR2 \\
\midrule

VR limitation
& Level of realism.
& Concerns that VR must reproduce meaningful lighting effects accurately enough to be pedagogically useful.
& Scope \\
\midrule

Instructor-interface needs: realism and operation
& Real fixture; safety training in rigging; cabling teaching; accessories demo; realistic hanging position; advanced color selection; detailed light settings and texture.
& Desired support for representing lighting equipment, operational procedures, and realistic lighting parameters.
& Scope, DR2 \\
\midrule

Instructor-interface needs: annotation and feedback
& Circle the area being discussed; different styled annotation for different categories; real-time tagging with voice chat and recognition; annotate within 3D environment and see instant change.
& Desired support for providing situated visual feedback and triggering immediate visual demonstrations during live teaching.
& DR1, DR2 \\
\midrule

Instructor-interface needs: review and documentation
& Comment history; whole session recording; voice capture and transcription; pinpoint specific phases; exportable/shareable recordings; documented feedback based on the environment.
& Desired support for capturing, reviewing, and sharing instructor feedback across time and scene contexts.
& DR3, Future work \\
\midrule

Instructor-interface needs: communication and collaboration
& Communication helper for professional terms; multi-user collaboration; peer feedback; breakout rooms; real-time file sharing.
& Desired support for terminology mediation, multi-stakeholder communication, and collaborative production-oriented learning.
& DR3, Future work \\
\midrule

Instructor-interface needs: performance and venue simulation
& Static and dynamic stages; venue simulation; different performance genres; detailed performer simulation; costumes on avatars.
& Desired support for teaching lighting design under different stage, performer, and performance conditions.
& Scope, Future work \\
\midrule

Real-time feedback rationale
& Instant feeling, emotion, and reaction; immediate guidance and correction; real-time comparison; precise feedback to dynamic activities.
& Reasons why real-time feedback is important for lighting education, especially when lighting effects and student activities change quickly.
& DR2 \\
\midrule

Feedback modality preferences
& AR; VR for beginner-level learning; computer plus face-to-face; VR for design and computer for paperwork; computer with consistent software.
& Preferences for feedback media depending on task type, accessibility, and instructor need to observe students.
& Scope \\
\midrule

Current instructional challenges
& Revision of student work; lack of real practice space; collaboration with production stakeholders; lack of instruments; lack of venue variety; limited attention; balancing momentum and feedback; varying skill levels.
& Constraints in current stage lighting instruction that motivate immersive, real-time, and learner-sensitive feedback systems.
& DR1--DR3 \\
\bottomrule
\end{tabular}
\end{table*}

\subsection{Prompts \& System Architecture Details}
\label{app:prompts_and_system_architecture}

\subsubsection{Meta-Prompt and Few-Shot Examples}
The core system prompt configures the AI assistant to act as a stage lighting course TA, instructing it to translate natural language teacher feedback into concrete light manipulation commands based on 32 numbered stage lights.

\begin{PromptBox}
You are a teaching assistant for a stage lighting course. Your responsibility is to convert the lighting suggestions input by the teacher into executable lighting actions for demonstration. You have 32 stage lights, with spotlightIds ranging from 1 to 32. From the audience's perspective:
8 top lights are positioned behind the central character on stage (slightly behind the character).
8 top lights are directly above the character.
10 front lights are placed in front of the character (illuminating the character's front), with these 26 lights (8 + 8 + 10) arranged centrally from left to right.
3 side lights are on the left and 3 on the right respectively: the left side lights illuminate the character's left side (from the audience's view), and the right side lights illuminate the character's right side, with each set of 3 arranged from front to back.

The 32 spotlightIds correspond to the following light descriptions in order:
1 - Back Top Light (Left 1)   ...   8 - Back Top Light (Left 8)
9 - Overhead Top Light (Left 1)   ...   16 - Overhead Top Light (Left 8)
17 - Front Light (Left 1)   ...   26 - Front Light (Left 10)
27 - Right Side Light (Front 1)   ...   29 - Right Side Light (Front 3)
30 - Left Side Light (Front 1)   ...   32 - Left Side Light (Front 3)

The actionable operations for each light are as follows:
- "on"
- "off"
- "highlight <object>" (Objects: "character", "white piano", "black piano")
- "direction line on" (include if rotation or aiming direction is implied)
- "color to <RGB>" (RGB format e.g., "color to 255,255,255". Never output "color to 0,0,0")
  Color-concept examples:
  * Opposite (complementary): ["color to 255,140,0", "color to 0,120,255"]
  * Adjacent: ["color to 255,80,120", "color to 255,170,80"]
  * Similar: ["color to 80,120,255", "color to 120,80,255"]
- "intensity to <value>" (0 to 100)
- "inner angle to <value>" (0 to current outer angle)
- "outer angle to <value>" (1 to 45)
- "rotate to" (re-aiming from one side toward another)

Important: All actions except "on", "off", and "highlight <object>" apply only to lights that are already on. If a light is off, output "on" first.
Ambiguity-rule: If using vague references like "this" or "these", default to operating on currently enabled lights.

Respond with ONLY a single JSON array of action objects.

--- FEW-SHOT EXAMPLES ---

### Dimension 1: Fixture & Configuration
Input: "Turn on the middle front light and disable light 5."
Output: [{"spotlight_id": "21", "action": "on"}, {"spotlight_id": "5", "action": "off"}]

### Dimension 2: Spatial Behavior
Input: "Re-aim the left side light toward the performer and tag frontlight on character."
Output: [
  {"spotlight_id": "30", "action": "rotate to"},
  {"spotlight_id": "30", "action": "direction line on"},
  {"spotlight_id": "character", "action": "highlight character"}
]

### Dimension 3: Visual Effects
Input: "Apply opposite color combination on the active side lights and widen the beam angle."
Output: [
  {"spotlight_id": "27", "action": "color to 255,140,0"},
  {"spotlight_id": "30", "action": "color to 0,120,255"},
  {"spotlight_id": "30", "action": "outer angle to 35"}
]

### Dimension 4: Composition Strategies
Input: "Create a lyrical and soft lighting atmosphere for a emotional solo performance."
Output: [
  {"spotlight_id": "21", "action": "on"},
  {"spotlight_id": "21", "action": "color to 255,200,150"},
  {"spotlight_id": "21", "action": "intensity to 40"},
  {"spotlight_id": "30", "action": "on"},
  {"spotlight_id": "30", "action": "color to 80,120,255"},
  {"spotlight_id": "30", "action": "intensity to 30"},
  {"spotlight_id": "character", "action": "highlight character"}
]
\end{PromptBox}

\subsubsection{Scene Metadata Serialization Format}
The real-time status of all active stage spotlights, including 3D spatial coordinates (`position`), orientation, cone angles, and color metrics, is serialized into JSON format and passed into the LLM context.

\begin{PromptBox}
### Current enabled lighting data (JSON payload passed into backend context)
[
  {
    "spotlight_id": "21",
    "enabled": true,
    "position": [0.0, 4.5, 3.2],
    "rotation": [45.0, 0.0, 0.0],
    "color": [255, 255, 255],
    "intensity": 80,
    "inner_angle": 15,
    "outer_angle": 30
  },
  {
    "spotlight_id": "30",
    "enabled": true,
    "position": [-3.5, 3.0, 1.0],
    "rotation": [30.0, 60.0, 0.0],
    "color": [0, 120, 255],
    "intensity": 60,
    "inner_angle": 10,
    "outer_angle": 25
  }
]
\end{PromptBox}

\subsubsection{Jargon Pipeline Prompt}
The Jargon Pipeline consists of two steps: extracting stage-lighting terms that literally appear in the teacher's transcript along with explanations, and mapping those terms to corresponding 3D scene positions.

\begin{PromptBox}
### Step 1: Jargon Extraction Prompt
You are a stage lighting glossary assistant. Given a transcript, extract stage-lighting-related terms (jargons) that **literally appear** in the transcript (exact wording or obvious plural/case variants). Do NOT infer or add synonyms that are not present. Skip generic words (e.g., 'light', 'direction').

Scope of terminology includes but not limited to:
- Stage areas/positions: Cyclorama, proscenium, apron, upstage, downstage, stage left/right, wings, grid, FOH.
- Fixture types/roles: frontlight, sidelight, backlight, toplight, followspot, ellipsoidal, fresnel, PAR, gobo, gel.
- Technical concepts: color temperature, color theory, complementary colors, beam/field angle, intensity, cue, fade.

Return ONLY JSON:
{
  "jargons": [
    {"name": "Frontlight", "explanation": "Lights the actor's face clearly from the front."}
  ]
}

### Step 2: Spatial Tag Mapping Prompt
You are a stage lighting expert. Given a transcript and a list of lighting jargons, decide which scene position each jargon is associated with.
Allowed positions only: character, cyclorama, apron, stage lip, stage right, stage left, upstage, downstage.

Rules:
1. Light types (front light, top light, key, wash) target performer -> 'character'.
2. Stage areas (cyclorama, apron, stage left) -> use position exactly.
3. Unmapped jargons -> 'none'.

Output JSON:
{"mappings": [{"jargon": "Frontlight", "position": "character"}]}
\end{PromptBox}

\subsection{Formative Study Materials}
\subsubsection{Expert Demographic Questionnaire}
\begin{enumerate}[leftmargin=*]
    \item \textbf{Age:} Below 20 / 20–30 / 30–40 / 40–50 / Above 50 / Prefer not to say
    \item \textbf{Gender:} Male / Female / Non-binary / Prefer not to say / Other
    \item \textbf{Nationality \& Country of Teaching:} (Open text)
    \item \textbf{VR Experience Level:} None / Basic / Intermediate / Advanced
    \item \textbf{Stage Lighting Teaching Experience:} (Years in field)
\end{enumerate}

\subsubsection{Semi-Structured Interview Protocol for Domain Experts}
Part 1: Current Practices and Tools
- How do you currently provide feedback to students in stage lighting courses?
- What tools, hardware, or software platforms do you use for instruction?

Part 2: Instructional Workflow
- Could you outline the typical process and key pedagogical milestones when instructing students in stage lighting design?

Part 3: VR Experiences and Suggestions
- Have you had any prior experience using VR for education? If so, describe the context.
- What potential advantages do you see in using VR for stage lighting instruction?
- What challenges or limitations have you encountered or do you anticipate when using VR?

Part 4: Interface Design Preferences
- If teaching in VR, what design requirements and factors should be considered?
- How important is real-time guidance and spatial feedback in your teaching approach?
- Would you prefer giving feedback directly within VR or via an external laptop/desktop interface?

Part 5: Challenges and Improvements
- What are the major bottlenecks in your current instructional methods regarding spatial and lighting concept delivery?

\subsection{User Study Guidance \& Scripts}
\subsubsection{Phase 1: Instructor Teaching Protocol \& Four Core Concepts}
\label{app:instructor_teaching_protocal}
All three instructors complete two 8–10 minute teaching sessions (Session A: Baseline/No LLM; Session B: LumiNote/With LLM) covering identical target concepts:
\begin{enumerate}[leftmargin=*]
    \item \textbf{Concept 1 (Side Light):} Definition and physical placement of side lighting.
    \item \textbf{Concept 2 (Color Dimensions):} Explanation of Hue, Saturation, and Brightness.
    \item \textbf{Concept 3 (Color Combinations):} Strategies for opposite (complementary), adjacent, and similar color pairs.
    \item \textbf{Concept 4 (Emotional Atmosphere):} Expressing mood through color (e.g., lyrical slow ballad vs. aggressive rock song).
\end{enumerate}

\subsubsection{Phase 2: Student Practical Tasks Script}
Students ($N=24$, between-subjects) watch a recorded instructor session in VR and perform four sequential lighting setup tasks:

Task 1 (Directional Side Light):
Add a directional side light pointing at the solo performer so that she clearly stands out from the surrounding stage background.

Task 2 (Light Quality Adjustment):
Increase the edge transition hardness and shadow sharpness of the side light to create a crisp, noticeable difference in lighting texture.

Task 3 (Color Contrast Strategy):
Apply a combination of warm and cool lighting colors across the active fixtures to establish clear visual color contrast in the scene.

Task 4 (Major Composition Task - Lyrical Performance):
Based on the current stage setup, create a complete lighting state suitable for a lyrical performance segment. The layout should demonstrate visual coherence in rhythm, color harmony, and spatial composition to evoke emotional immersion.

\subsection{Evaluation Questionnaires \& Rubrics}
\subsubsection{Instructor Evaluation Questionnaires \& Interview Protocol}

\paragraph{Quantitative Survey Scales}
Table~\ref{tab:instructor_questionnaire} outlines the custom and standardized items used for the instructor evaluation.

\paragraph{Instructor Post-Study Interview Protocol}
\begin{enumerate}[leftmargin=*]
    \item \textbf{Communication Burden:} Compared with your traditional teaching methods, did LumiNote reduce your communication cost and conceptual explanation burden when teaching lighting direction and color concepts? If so, in what specific ways?
    \item \textbf{Session Preference:} Which session (No LLM vs. With LLM) did you prefer, and why?
    \item \textbf{Feature Utility:} Which functions of LumiNote—including LLM-generated annotations, multimodal interaction, and spatial visualizations—did you find most helpful for supporting real-time instruction, and why?
    \item \textbf{VR Adoption Willingness:} Would you be more willing to use VR for teaching with the assistance of this system? To what extent?
    \item \textbf{Teaching Prompts \& Reminders:} Did the system ever help remind you to explain something you might have otherwise missed or forgotten? (e.g., providing additional explanations that prompted you to give more instructions)
    \item \textbf{Vision \& Future Improvements:} Beyond the current design, what additional capabilities or interaction patterns would you envision for an LLM-assisted VR lighting teaching system to better support professional stage lighting education?
\end{enumerate}

\subsubsection{Student Evaluation Questionnaires \& Interview Protocol}

\paragraph{Quantitative Survey Scales}
Table~\ref{tab:student_questionnaire} details the custom and standardized items evaluated by student participants.

\paragraph{Student Post-Study Interview Protocol}
\begin{enumerate}[leftmargin=*]
    \item \textbf{Effective Learning Factors:} Which part or element of the teaching session helped you learn best?
    \item \textbf{Clarity \& Gap Identification:} Was there anything you wished the teaching session had shown more clearly, or something that could have helped you understand better?
    \item \textbf{Task Guidance Utility:} When you were doing the tasks after watching the teaching session, what helped you understand what you needed to do?
    \item \textbf{Learning Experience Improvement:} If you could change one thing about this learning experience to make it better, what would it be?
\end{enumerate}

\subsubsection{Expert Grading Rubric for Student Major Task}
Student performance in the major practical task (Task 4: Designing a lighting setup for a lyrical performance) is evaluated asynchronously by expert instructors. The assessment is based on a **paired submission**: a high-resolution screenshot of the student's finalized VR lighting scene accompanied by their transcribed rationale explaining their design choices and underlying concepts.

\subsection{Use of Artificial Intelligence}
The authors used generative AI tools to assist with manuscript writing and code development. All AI-assisted writing and code were reviewed and revised by the authors, and all code and reported results were manually double-checked for correctness. The authors take full responsibility for the final content of this work.

\clearpage
\begin{table*}[!htbp]
\caption{Instructor Quantitative Evaluation Questionnaire}
\label{tab:instructor_questionnaire}
\small
\begin{tabular}{lp{5.8cm}r}
\toprule
\textbf{Category / Construct} & \textbf{Item Text / Scope} & \textbf{Scale} \\
\midrule
\textbf{Feature Support} & AI-generated lighting demonstrations & 1--7 \\
 & Green laser pointer as direction indicator & 1--7 \\
 & Spatial arrow annotations for lighting direction & 1--7 \\
 & Star-shaped highlights in scene & 1--7 \\
 & Lighting concept text tagging in scene & 1--7 \\
 & Tagging on lighting console panel with actions and suggested values & 1--7 \\
 & Transcription panel with jargon highlighted and explained & 1--7 \\
 & Animated instructor avatar & 1--7 \\
\midrule
\textbf{Instructional Quality} & I found it easy to explain lighting concepts effectively with this system. & 1--7 \\
\textbf{\& Effectiveness} & I found it easy to give instructions and guidance clearly with this system. & 1--7 \\
 & The system helps me teach with professionalism. & 1--7 \\
 & The system helps me present content in a structured and expert manner. & 1--7 \\
 & The system responded quickly to actions and inputs during the session. & 1--7 \\
 & The system supported real-time interaction without noticeable delay. & 1--7 \\
 & The recorded session is effective at helping students understand abstract concepts. & 1--7 \\
 & The recorded session is effective at improving students' ability to apply concepts. & 1--7 \\
 & The content created with the system was intuitive for understanding the concept. & 1--7 \\
 & The content within the system was clear and easy to understand. & 1--7 \\
\midrule
\textbf{Standard Measures} & NASA-TLX Workload (Mental, Physical, Temporal, Performance, Effort, Frustration) & 1--7 \\
 & IPQ Presence (Spatial Presence, Realism, Naturalness, Involvement; 8 items) & 1--7 \\
 & System Usability Scale (SUS; standard 10 items) & 1--5 \\
\bottomrule
\end{tabular}
\end{table*}

\begin{table*}[!htbp]
\centering
\caption{Student Quantitative Evaluation Questionnaire}
\label{tab:student_questionnaire}
\small
\begin{tabular}{lp{5.8cm}r}
\toprule
\textbf{Category / Construct} & \textbf{Item Text / Scope} & \textbf{Scale} \\
\midrule
\textbf{Prior Knowledge} & Before this study, how familiar are you with: & \\
 & -- Lighting Directionality and Subject Definition & 1--5 \\
 & -- Light Quality and Visual Appearance & 1--5 \\
 & -- Color Relationships in Lighting and Scene Composition & 1--5 \\
\midrule
\textbf{Feature Support} & 8 LumiNote features (Demos, Laser, Arrows, Star Highlights, Tags, Console Tags, Jargon Panel, Avatar) & 1--7 \\
\midrule
\textbf{Learning Perception} & The recorded session is effective at helping me understand abstract lighting concepts. & 1--7 \\
\textbf{\& Self-Confidence} & The recorded session is effective at improving my ability to apply learned concepts. & 1--7 \\
 & The content within the system was intuitive for understanding the teaching concept. & 1--7 \\
 & The content within the system was clear and easy to understand. & 1--7 \\
 & I feel confident that I understood the lighting concept taught during the session. & 1--7 \\
 & I feel confident that I completed the tasks correctly. & 1--7 \\
\midrule
\textbf{Standard Measures} & NASA-TLX Workload (6 standard items) & 1--7 \\
 & IPQ Presence Inventory (8 standard items) & 1--7 \\
 & System Usability Scale (SUS; standard 10 items) & 1--5 \\
\bottomrule
\end{tabular}
\end{table*}

\begin{table*}[!htbp]
\centering
\caption{Expert Evaluation Rubric for Student Major Task Artwork \& Rationale}
\label{tab:student_major_task_rubric}
\small
\begin{tabular}{lp{5.2cm}r}
\toprule
\textbf{Dimension} & \textbf{Evaluation Criteria} & \textbf{Weight} \\
\midrule
\textbf{Concept Accuracy} & Accuracy and depth in applying taught lighting concepts (e.g., key/side lighting roles, color temperature contrast, beam sharpness) as articulated in the student's rationale. & 35\% \\
\textbf{Visual Atmosphere} & Aesthetic quality and emotional alignment of the rendered scene screenshot (e.g., mood coherence, spatial composition, harmony with a lyrical ballad). & 35\% \\
\textbf{Concept Comprehensiveness} & Extent to which multiple lighting dimensions (Direction, Light Quality, and Color Relationships) are holistically integrated in both the visual scene and justification. & 30\% \\
\bottomrule
\end{tabular}
\end{table*}









\end{document}